\documentclass[useAMS,usenatbib]{mn2e}
\usepackage{graphicx}
\usepackage{color}
\usepackage{multirow}

\newcommand\Msun{M$_\odot$}

\newcommand\ha{H\,{$\alpha$}}

\newcommand\oiii{[O\,{\sevensize III}]}
\newcommand\oii{[O\,{\sevensize II}]}
\newcommand\nii{[N\,{\sevensize II}]}

\newcommand\NBH{NB$_{\rm H}$}
\newcommand\apj{ApJ}

\newcommand\apjl{ApJL}
\newcommand\mnras{MNRAS}
\newcommand\aap{A\&A}
\newcommand\aj{AJ}
\newcommand\pasj{PASJ}
\newcommand\pasp{PASP}

\title[3-D structures at $z=1.46$]{
Mapping the large scale structure around a z=1.46 galaxy cluster in
3-D using two adjacent narrow-band filters  
} 
\author[M. Hayashi et al.]{%
Masao Hayashi$^{1}$\thanks{E-mail: hayashim@icrr.u-tokyo.ac.jp},
Tadayuki Kodama$^{2,4}$,
Yusei Koyama$^{2}$,
Ken-ichi Tadaki$^{2}$,
\newauthor
Ichi Tanaka$^{3}$,
Rhythm Shimakawa$^{3,4}$, 
Yuichi Matsuda$^{4,5}$, David Sobral$^{6,7}$, 
\newauthor Philip N.~Best$^{8}$, Ian Smail$^{9}$\\
$^{1}$Institute for Cosmic Ray Research, The University of Tokyo, Kashiwa, Chiba 277-8582, Japan\\  
$^{2}$Optical and Infrared Astronomy Division, National Astronomical Observatory, Mitaka, Tokyo 181-8588, Japan\\
$^{3}$Subaru Telescope, National Astronomical Observatory of Japan, 650 North A'ohoku Place, Hilo, HI 96720, USA\\
$^{4}$Department of Astronomical Science, The Graduate University for
Advanced Studies (SOKENDAI), Mitaka, Tokyo 181-8588, Japan\\
$^{5}$Radio Astronomy Division, National Astronomical Observatory of Japan, 2-21-1 Osawa, Mitaka, Tokyo 181-8588, Japan\\
$^{6}$Leiden Observatory, Leiden University, P.O. Box 9513, NL-2300 RA Leiden, THe Netherlands\\
$^{7}$Centro de Astronomia e Astrof\'{\i}sica da Universidade de Lisboa, Observat\'{o}rio Astron\'{o}mico de Lisboa,\\ 
\ Tapada da Ajuda, 1349-018 Lisboa, Portugal\\
$^{8}$SUPA, Institute for Astronomy, Royal Observatory of Edinburgh, Blackford Hill, Edinburgh, EH9 3HJ, UK\\
$^{9}$Institute for Computational Cosmology, Durham University, South Road, Durham, DH1 3LE, UK\\
}

\begin{document}

\date{Accepted 2014 January 15.  Received 2014 January 7; in original form 2013 November 7}

\pagerange{\pageref{firstpage}--\pageref{lastpage}} \pubyear{2014}

\maketitle

\label{firstpage}

\begin{abstract}
We present a novel method to estimate accurate redshifts of
star-forming galaxies by measuring the flux ratio of the same emission
line observed through two adjacent narrow-band filters. 
We apply this method to our NB912 and new NB921 data taken with
Suprime-Cam on the Subaru Telescope of a galaxy cluster,
XMMXCS~J2215.9-1738, at $z=1.46$ and its surrounding structures. 
We obtain redshifts for 170 \oii\ emission line galaxies at
$z\sim1.46$, among which 41 galaxies are spectroscopically confirmed
with MOIRCS and FMOS on the Subaru mainly, showing an accuracy of
$\sigma((z-z_{\rm spec})/(1+z_{\rm spec}))=0.002$.   
This allows us to reveal filamentary structures that penetrate towards
the centre of the galaxy cluster and intersect with other structures,
consistent with the picture of hierarchical cluster formation. 
We also find that the projected celestial distribution does not
precisely trace the real distribution of galaxies, indicating the
importance of the three dimensional view of structures to properly
identify and quantify galaxy environments. We investigate the
environmental dependence of galaxy properties with local density,
confirming that the median colour of galaxies becomes redder in higher
density region while the star-formation rate of star-forming 
galaxies does not depend strongly on local environment in this
structure. This implies that the star-forming activity in galaxies is
truncated on a relatively short time scale in the cluster centre.   
\end{abstract}

\begin{keywords}
galaxies: clusters: general -- galaxies: evolution.
\end{keywords}

\section{Introduction}
\label{sec;introduction}

Galaxies are distributed inhomogeneously on small scales in the
Universe and thus define large-scale structures which consist of dense
regions of galaxies such as galaxy clusters, sparse regions called
voids, and filamentary structures which connect the dense regions, as
revealed by large spectroscopic surveys such as Sloan Digital Sky
Survey \citep[SDSS,][]{York2000} and 2 degree Field Galaxy Redshift
Survey \citep[2dFGRS,][]{Colless2001}. Galaxy clusters embedded in the
large-scale structures are some of the densest systems in the Universe 
and occur at intersections of the galaxy filaments. While the spatial
distribution of galaxies has been revealed in detail in the local
Universe, the process by which the galaxy clusters form is yet to be
understood in detail.

Galaxy clusters are considered to originate from the clustering of
some young galaxies (i.e., proto-cluster) in the early Universe and to
gradually grow into more massive gravitationally bound systems
with time while accreting galaxies from the surrounding regions. The
standard cold dark matter (CDM) scenario can reproduce well the 
picture seen in the Universe within the framework of the hierarchical
formation of galaxy structures \citep[e.g.,][]{Springel2005}.  
Observational studies through wide-field imaging surveys of distant
clusters at $z<1$ have also supported the hierarchical formation by
finding large-scale structures around galaxy clusters 
\citep[e.g.,][]{kodama2005, gal2008, tanaka2009, koyama2010}.        
Now, the discovery of large-scale structures around galaxy clusters and
proto-clusters is being extended to $z\sim3$  
\citep[e.g.,][]{tanaka.i2011,tadaki2012,hayashi2012,koyama2013a,Erb2011,Yamada2012,Galametz2013,Henry2014}. 
One interesting finding suggested by such observational studies is that
cluster galaxies seem to follow an inside-out evolution in the sense
that the environment where galaxies are active shifts from the densest
regions at high redshifts ($z\approx$ 2--3) towards less dense regions
at lower redshifts ($z\ll1$). 
Star forming activity increases rapidly in the denser regions and becomes
comparable to the surrounding lower density regions at higher
redshifts, which implies that galaxy formation is biased towards
higher density regions at high redshift \citep[][and references
therein]{kodama2013}. However, it is also known that the environmental
dependences of galaxy properties seen at $z\sim0$ are already in place
at $z\sim1$ \citep[e.g.,][]{Holden2007,vanderWel2007,Patel2009,Patel2011,Sobral2011}.  

The next step to understand the evolution of cluster galaxies is to
determine precisely how galaxies change their properties as a result
of the hierarchical growth of large-scale structures.  
However, the difficulty is the need for accurate determination of
the redshift to determine the precise environment of the
galaxies. Although photometric redshifts estimated by fitting template
spectral energy distributions (SEDs) to the multi-wavelength
photometry of galaxies is quite effective, the uncertainty becomes
large at higher redshifts. Moreover, the number density of blue
star-forming galaxies with less prominent features in the SED exceeds
that of red quiescent galaxies which have strong breaks in their SED
at $z \ga$1.0--1.5 \citep[e.g.,][]{Brammer2011,Moustakas2013,Muzzin2013,Tomczak2013}.
An effective way to solve the issue is by carrying out the imaging with
narrowband filters targeting nebular emission from HII regions of
star-forming galaxies. The narrowband imaging enables a largely
unbiased sampling of star-forming galaxies down to a limiting star-formation
rate of galaxy and the redshift of galaxies selected are determined  
with the accuracy of $\Delta z \sim0.03$.   
However, this is too uncertain to accurately define the environment of
galaxies as this redshift range corresponds to a co-moving depth of
$\sim$ 56 Mpc at $z=1.46$. We thus demonstrate a novel technique to
circumvent this problem.

The galaxy cluster XMMXCS~J2215.9-1738 (hereafter XCS2215) at $z=1.46$
is one of the most distant massive clusters known \citep{stanford2006}. 
Extended X-ray emission from hot gas bound in the cluster suggests
that it is already a mature system. However, previous studies have
revealed that this galaxy cluster is still growing actively
\citep{hayashi2010,hayashi2011}. 
We discovered a high fraction of galaxies undergoing active star
formation in the core of the cluster and the existence of large scale
structures of star-forming galaxies around the cluster. The finding
means that this galaxy cluster is a good target to investigate the
processes operating on galaxies assembling into the centre of the
cluster from the outskirts. However, our current understanding is
based on a view of the large scale structures projected on the
celestial plane. We require to know the true three dimensional
structures to reveal in detail the processes playing a critical role
in formation of characteristic properties of cluster galaxies, because
the projected distribution of galaxies may not reflect the real
structures. The results found without the three dimensional structures
can be led to wrong conclusions. Thus, in this paper, we aim to
define the three dimensional structures around the XCS2215 cluster at
$z=1.46$. 

The outline of this paper is as follows. The observations and data
are described in \S~\ref{sec;data}, and 
samples of emission line galaxies identified with three different 
narrowband filters are presented in \S~\ref{sec;catalogues}.    
A new method to estimate the accurate redshifts with two adjacent
narrowband filters is introduced in \S~\ref{sec;redshifts}, which
allows us to reveal the three dimensional view of the filamentary
large-scale structures around the galaxy cluster in
\S~\ref{sec;results}. Then, we compare \oii\ emission line galaxies
with \ha\ emission line galaxies associated with these structures. We
also discuss the dependence of galaxy properties on the environment
defined by the three dimensional distribution of the galaxies. 
Finally, we present the conclusions of this paper in \S~\ref{sec;conclusions}.   
Throughout this paper, magnitudes are presented in the AB
system. However, Vega magnitudes in $J$ and $K$, if preferred, can be
obtained from our AB magnitudes using the relations: $J_{\rm Vega}$ =
$J_{\rm AB}-0.92$ and $K_{\rm Vega}$ = $K_{\rm AB}-1.90$, respectively.  
We adopt cosmological parameters of $h=0.7$, $\Omega_{m}=0.3$ and
$\Omega_{\Lambda}=0.7$. In $z$ = 1.46, 1 arcmin corresponds to 1.25 Mpc
(co-moving) and 0.51 Mpc (physical), respectively. 

\section{Observations and Data}
\label{sec;data}

\subsection{Narrowband filters targeting nebular emission lines at $z$=1.46}
\begin{figure}
 \begin{center}
 \includegraphics[width=\linewidth]{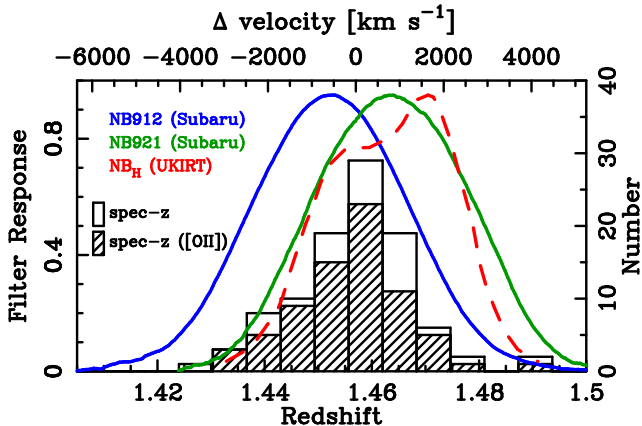}
 \end{center}
 \caption{ Response curves of NB912 (blue), NB921 (green) and \NBH\
   (red) filters. The horizontal axis shows the redshifts where \oii\
   or \ha\ lines enter each filter; \oii\ for NB912 and NB921, and
   \ha\ for \NBH. The open and hatched histograms show the distribution
   of spectroscopic redshift of all cluster members and \oii\ emitters
   selected with NB912 and NB921, respectively. The \oii\ emitters are
   spectroscopically confirmed using MOIRCS \citep{hayashi2011} and
   FMOS (\S~\ref{sec;specdata}) on the Subaru Telescope. 
  }   
 \label{fig;NBfilters}
\end{figure}

We used two narrowband filters which are adjacent filters and both can
detect \oii\ emission at $z\sim1.46$. 
These are the NB912 and NB921 narrowband filters installed on the
Subaru Prime Focus Camera \citep[Suprime-Cam;][]{miyazaki2002} on the
Subaru Telescope.  
The central wavelength and full width half maximum (FWHM) of the NB912
(NB921) filters are 9139 (9196)\AA\ and 134(132)\AA, which implies a
velocity difference of $\sim2000$ km s$^{-1}$ for galaxies at $z=1.46$
(Figure \ref{fig;NBfilters}). The specifications of the two
narrow-band filters are shown in Table~\ref{tbl;NBfilters}. 
Our analysis also makes use of a third filter: the \NBH\ narrowband
filter installed on the Wide Field Camera \citep[WFCAM;][]{casali2007} on
the the United Kingdom Infrared Telescope (UKIRT). This filter was
made for the High Redshift Emission Line Survey (HiZELS) which is a
narrowband imaging survey aiming to detect \ha\ emission from
galaxies at $z=$0.4, 0.8, 1.5, and 2.2 with WFCAM on the UKIRT
\citep{Geach2008,Sobral2009,Sobral2012,Sobral2013,Best2010},  
and can detect \ha\ emission at $z\sim1.46$ with the central
wavelength and FWHM of response curve being 1.617\micron\ and
0.0211\micron. As shown in Figure \ref{fig;NBfilters}, the response
curves of the NB921 and \NBH\ filters are well matched in redshift
space corresponding to \oii\ or \ha\ emission lines at $z\sim1.46$
\citep{Sobral2012,hayashi2013}, while the NB912 filter is suitable to
select \oii\ emitters slightly in the foreground.    

\begin{table*}
 \caption{Specifications of the three narrowband filters and the
   samples of emission line galaxies.}
\begin{center}
\begin{tabular}{ccccccccc}
\hline\hline
Filter & $\lambda_c$ & $\Delta\lambda$ & Telescope &
\multicolumn{2}{c}{Emitter}& \multicolumn{2}{c}{NB912+NB921 \oii} &Limiting SFR$^\ddagger$ \\
 &  [\micron] & [\micron] &  & Line & Number & Number & Spec.~confirmed$^\dagger$ &[\Msun/yr] \\
\hline
NB912 & 0.9139 & 0.0134 & Subaru & \oii & 380 & \multirow{2}{*}{170} & \multirow{2}{*}{41} & 2.6\\
NB921 & 0.9196 & 0.0132 & Subaru & \oii & 429 &  &  &  2.2\\
\NBH   & 1.617 & 0.0211 & UKIRT  & \ha  &   9 &  1  & -- & 14.6\\
\hline\hline
\multicolumn{9}{l}{$\dagger$ We spectroscopically confirmed 73 \oii\
  emitters in total; NB912+NB921=41, NB912=21, and NB921=11.}\\
\multicolumn{9}{l}{$\ddagger$ The limiting SFRs are not corrected for dust extinction.}
\label{tbl;NBfilters}
\end{tabular}
\end{center}
\end{table*}

\subsection{The NB921 data}

The NB921 data were obtained with Suprime-Cam on the Subaru Telescope,
which has a field-of-view (FoV) of 34$\times$27 arcmin$^2$,
corresponding to 42.5$\times$33.8 Mpc$^2$ in co-moving scale at
$z=1.46$, under the open-use programme in service-mode (S12A-131S, PI:
M. Hayashi) on 2012 July 24. This enables us to cover from the central
region to the outskirts of the cluster with a single pointing.  
We took 11 frames with individual exposure time of 20 minutes, which
results in the total integration time of 220 minutes. The observations
were conducted under photometric conditions. The seeing varied between
0.6 and 0.8 arcsec.    

We used the data reduction package for Suprime-Cam 
\citep[SDFRED2:][]{yagi2002,ouchi2004} 
to reduce the data in the same way as for the other
optical data \citep{hayashi2010,hayashi2011}.
Note that we also used the software released by \citet{yagi2012} to
correct for the crosstalk seen around bright stars in the Suprime-Cam
image. The point spread function (PSF) is matched to 1.09 arcsec,
which is the PSF of the other optical data (see
\S~\ref{sec;opticaldata}). The photometric zero-point is determined
using the spectrophotometric standard star, Feige110. The 5$\sigma$
limiting magnitude measured within a 2\arcsec-diameter aperture is
25.4 comparable to the depth of 25.2 in the existing NB912 data (see
\S~\ref{sec;opticaldata}).    

\subsection{The \NBH\ and H data from HiZELS}
\label{sec;HiZELSdata}

As a part of HiZELS, we conducted the \NBH\ narrowband imaging in the
galaxy cluster. WFCAM consists of four 2048$\times$2048 HgCdTe
detectors with a pixel scale of 0.4\arcsec, and each detector covers a
13.65\arcmin$\times$13.65\arcmin\ region. However, the detectors are   
spaced with a gap of 12.83\arcmin. Thus, we observed the galaxy
cluster in \NBH\ and $H$-bands by pointing one of the detector at the
centre of cluster, resulting in the central
13.65\arcmin$\times$13.65\arcmin\ images being available in both bands.   
Note that the broadband image is required to estimate the continuum
flux density underneath the \ha\ emission line.

Data reduction was conducted in the standard procedure with the
pipeline for HiZELS \citep[PfHiZELS:][]{Geach2008,Sobral2009}. The
integration times are 256.7 and 49 minutes in \NBH\ and $H$,
respectively. Since seeing sizes were 1.02 and 1.32 arcsec in \NBH\
and $H$, they are matched to 1.32 arcsec. The 5$\sigma$ limiting
magnitudes are 21.61 in \NBH\ and 22.28 in $H$, which are measured
with a 2.4\arcsec-diameter aperture. 

\subsection{The data set used for previous studies}
\label{sec;opticaldata}
Our previous studies in the galaxy cluster are already published in
\citet{hayashi2010,hayashi2011} using six broadband images in $B, R_c,
i', z', J$ and $K$ and the NB912 narrowband image.  
Thus, we briefly summarize these data here. 

Deep, wide-field imaging data in optical were taken with Suprime-Cam
on the Subaru Telescope. The PSFs in the optical images are all
matched to 1.09 arcsec. The 5$\sigma$ limiting magnitudes are between
25.2 and 27.0 \citep{hayashi2011}.     
The data set in the near-infrared consists of $J$ and $K$-band images.
$K$-band data were taken with WFCAM on the UKIRT and overlap well with
the optical images. However, $J$-band images taken with 
Multi-Object Infrared Camera and Spectrograph
\citep[MOIRCS;][]{ichikawa2006,suzuki2008} on the Subaru telescope are
available in the central 6\arcmin$\times$6\arcmin\ region only. PSFs
in both $J$ and $K$ are also matched to 1.09 arcsec. The 5$\sigma$
limiting magnitudes are 23.2 and 22.7 in $J$ and $K$, respectively,
which are measured with a 2\arcsec-diameter aperture
\citep{hayashi2011}. 

\subsection{Photometric catalogue}

In this section, we describe the two catalogues for NB921-detected and
\NBH-detected sources. The NB912-detected catalogue is fully described
in \citet{hayashi2010,hayashi2011}.

\subsubsection{NB921-detected catalogue}

A photometric catalogue for NB921-detected sources is made following
\citet{hayashi2010, hayashi2011}. 
We use {\sc SExtractor} \citep[ver. 2.8.6:][]{bertin1996} to detect
sources on the NB921 image. Note that we use the NB921 image before
the PSF matching (i.e.~FWHM of 0.81\arcsec) for the source
detection to take advantage of the better signal-to-noise
ratio. Photometry is conducted on all the PSF-matched 
images by using {\sc SExtractor} in dual-image mode. 
Colours are derived from the 2\arcsec-diameter aperture
magnitudes, and {\sc mag\_auto} magnitudes are used as total
magnitudes. 
The aperture size is determined so as to be about twice as large as the PSF.
Magnitude errors are estimated from the 1$\sigma$ sky noise
taking account of the difference in depth at each object position due
to slightly different exposure times and sensitivities.
Magnitudes are corrected for the Galactic absorption by the following
magnitudes; A($B$)=0.10, A($R_c$)=0.06, A($i'$)=0.05, A($z'$)=0.04,
A(NB921)=0.04, A($J$)=0.02, and A($K$)=0.01, which are derived from
the extinction law of \citet{cardelli1989} on the assumption of
$R_V=3.1$ and $E(B-V)=0.025$ estimated from \citet{schlegel1998}.
The values for the correction are the same as used for the
NB912-detected catalogue of \citet{hayashi2011}.
To check the NB921 zero-point, we compare NB921 magnitudes with NB912
magnitudes for stellar sources. We find that there is a good agreement
between both magnitudes, indicating the validity of zero-point of
magnitude in NB921 image. 

As a result, after rejecting objects in the regions masked due to
the bad quality of the images, the catalogue contains 40,237 sources
brighter than 25.44 mag in NB921 (5$\sigma$ limiting
magnitude). Among them, 36,415 galaxies are distinguished from 3,822
stars based on $B-z'$ and $z'-K$ colours. This technique was devised
by \citet{daddi2004}, and stars are actually well separated from
galaxies on this colour-colour diagram ($B-z'$ vs.\ $z'-K$)
\citep{daddi2004,kong2006}.  

\subsubsection{\NBH-detected catalogue}

A catalogue for \NBH-detected sources is made by following the steps
described above. Since the PSF size of the \NBH\ and $H$ images are
larger than that of the others, this catalogue is used separately to
select the \NBH\ \ha\ emitters at $z\sim1.46$ in \S~\ref{sec;catalogues}.  
The \NBH\ image with PSF of 1.02\arcsec\ is used for source detection,
and photometry is conducted on the PSF-matched images using the
dual-image mode of {\sc SExtractor}. The $H$--\NBH\ colour is obtained
from 2.4\arcsec-diameter aperture magnitudes. Magnitudes are corrected
for the Galactic absorption by A(\NBH)=0.01 and A($H$)=0.01.
To compile the photometry for all of the data except $H$ and \NBH, 
we cross-match the \NBH-detected sources with those detected by NB912
or NB921. For almost all \NBH-detected sources, counterparts are found
in the NB912 or NB921 catalogues. We note that the \NBH-detected
sources for which no counterpart is found are cross-talk and false
detection seen around bright stars. Thus, we remove them from the
catalogue. As a result, the catalogue contains 1,324 sources brighter
than 21.61 mag. in \NBH\ (5$\sigma$ limiting magnitude).

\begin{figure*}
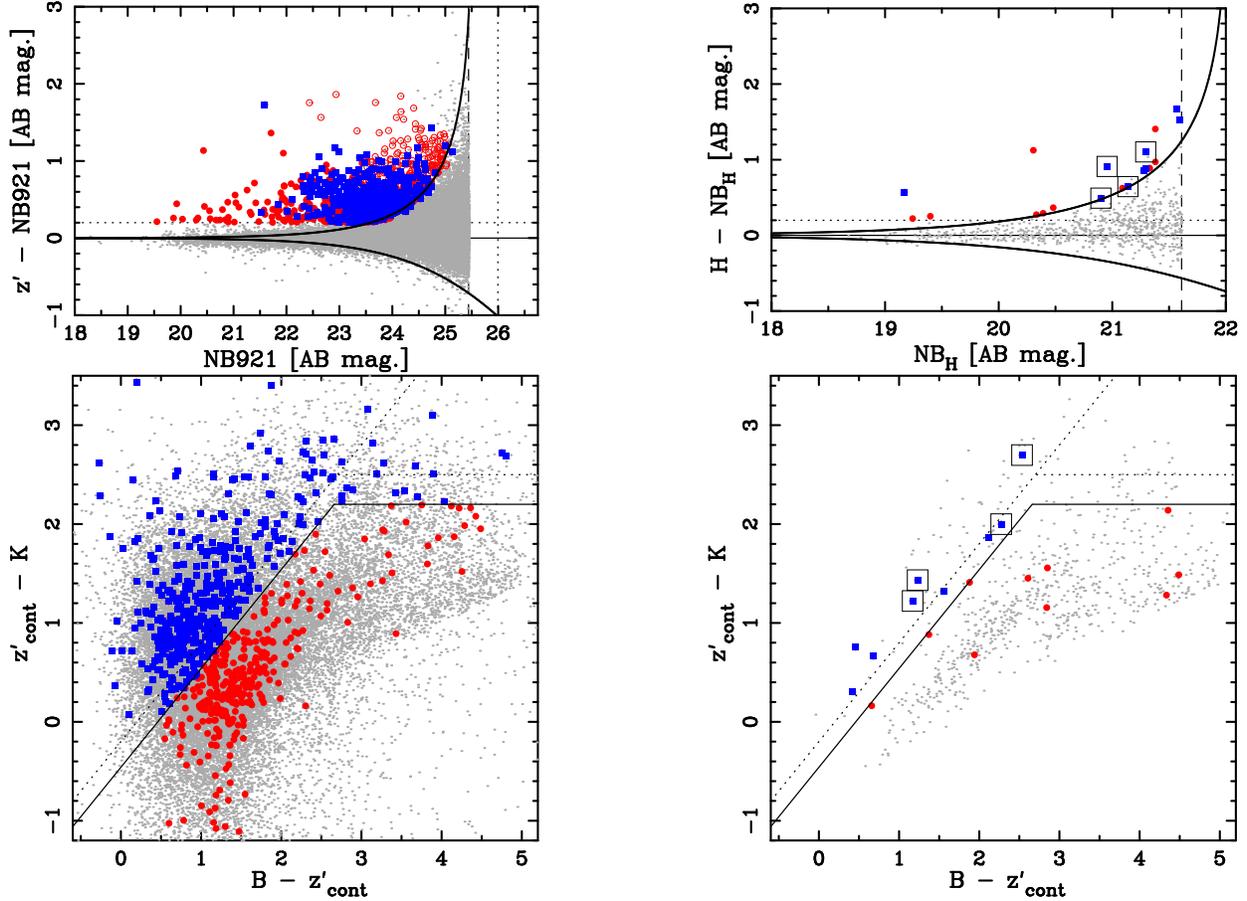

    \begin{tabular}{cc}

      \begin{minipage}{0.5\linewidth}
        \begin{center}
          \includegraphics[width=0.8\linewidth]{./fig2a.eps}
        \end{center}
      \end{minipage} &
      \begin{minipage}{0.5\linewidth}
        \begin{center}
          \includegraphics[width=0.8\linewidth]{./fig2b.eps} 
        \end{center}
      \end{minipage}\\

      \begin{minipage}{0.5\linewidth}
        \begin{center}
          \includegraphics[width=0.8\linewidth]{./fig2c.eps} 
        \end{center}
      \end{minipage} &
      \begin{minipage}{0.5\linewidth}
        \begin{center}
          \includegraphics[width=0.8\linewidth]{./fig2d.eps}
        \end{center}
      \end{minipage} 

    \end{tabular}
  \caption{ 
    Top: Broadband - Narrowband colours as a function of Narrowband
    magnitude (aperture magnitude). Bottom: two colour diagrams used to
    select emission line galaxies at $z=1.46$. In the left two panels
    for NB921 \oii\ emitters, emission line galaxies (un-)detected in
    $K$-band are shown by (open) filled red circles and then
    \oii\ emitters identified are shown by blue filled squares. Gray
    dots are all galaxies in the NB921-detected catalogue. In the left
    lower panel, solid black lines are our colour selection criteria
    defined in \citet{hayashi2011}, while dotted lines shows the
    original criteria for the selection of $BzK$ galaxies
    \citep{daddi2004}. The right two panels are the same as the left
    ones, but for \NBH\ \ha\ emitters. Blue symbols marked with open
    squares show \ha\ emitters having a detection of \oii\ emission in
    NB912 or NB921 data. 
  }   
  \label{fig;selection}
\end{figure*}

\subsection{Spectroscopic data}
\label{sec;specdata}

\citet{hayashi2011} reported the confirmation of sixteen NB912 \oii\
emitters in the central region of the galaxy cluster using MOIRCS on
the Subaru Telescope. These data were useful for showing the 
validity of our selection of NB912 \oii\ emitters. However, the
spectroscopic confirmation was limited to \oii\ emitters in the
core region.    

We have used the Fiber Multi Object Spectrograph (FMOS) on the Subaru
Telescope to confirm the large scale structure of \oii\ emitters we
have discovered in \citet{hayashi2011}. It allows for multi-object
spectroscopy in the $J$ and $H$-bands with 400 fibers and a wide FoV of 30
arcmin. The FMOS spectroscopy was performed under an open-use program
(S12A-084, PI: T. Kodama) during one and half nights of 2012 June 22--23. 
Unfortunately, due to the instrumental trouble, only the IRS2, i.e.,
one of the two spectrographs in FMOS, was available in our observation
run. Therefore, only 50\% of sources were targeted, resulting in
sparse sampling of the targets. Targets are mainly selected from the
sample of NB912 \oii\ emitters shown in \citet{hayashi2011}. We
targetted 78 galaxies in total, of which 56 are NB912 \oii\ emitters. 
In this run, we used the high resolution H-long setting which covers
the wavelength of 1.6--1.8\micron\ with a spectral resolution of
R=2000, enough to resolve \ha\ and \nii\ emission lines, where present. 
We adopted the cross beam switching method, that is, half of the
fibers available are allocated to the targets and at the same time the
other half of them observe nearby regions of blank sky. The total
integration time was 5.5 hours. Observations were conducted in
photometric conditions. The seeing was 0.7--0.9 arcsec in $R$-band,
which was measured for the coordinate calibration stars (so-called CCS
stars in the FMOS spine-to-object allocation software) near the centre of
the FoV, implying that loss of flux escaping from the 1.2-arcsec
diameter fibre should be small for unresolved objects. 

We had another opportunity to observe the \oii\ emitters with FMOS for
about two hours on 2013 October 13 (S13B-106, PI: K.-I.~Tadaki). The
targets are selected from the samples of NB912 \oii\ emitters and
NB921 \oii\ emitters described below. The same setup as for the
previous observation was used in this run. However, only one
spectrograph (IRS1) was available again. The total 
integration time was 1.5 hours on source. During the observations, the
sky was almost clear and the seeing was $\sim$1.0 arcsec. Spectra were
obtained for 65 \oii\ emitters. 

We used the FIBRE-pac data reduction package \citep{iwamuro2012} to
reduce the FMOS data. FIBRE-pac corrects the flux for the loss from
the fiber which is estimated using a calibration star.    
Although further correction may be required for extended sources like
galaxies \citep{yabe2012,Stott2013}, the redshift and ratio of emission line
fluxes can be measured reliably even if the correction is not
performed. We fitted three Gaussians to the spectra to measure \ha\
and the \nii\ doublet and estimated the signal-to-noise ratio of
emission lines seen in the spectra, and the redshift. In the Gaussian
fitting, the width of the two Gaussians fitted to the \nii\ doublet
is fixed to the same value as \ha. In addition, the flux ratio of \nii\
doublet is fixed to one-third. 
Errors on the emission line fluxes are estimated using the standard
deviation of 500 measurements of the flux, where in each measurement a
set of three Gaussians is fitted to the spectrum to which the error
with normal distribution with 1$\sigma$ spectral noise is added. 
We regard \ha\ emission with flux greater than 3$\sigma$ as a
detection, and then check the spectra judged to have a line
detection, by visual inspection.  
We succeed in spectroscopically confirming 48 sources by
detecting \ha\ emission at more than 3$\sigma$ in the whole region we
surveyed, among which the number of the \oii\ emitters is 41.  
Even if only a single emission line is detected in the spectrum, it
can be identified as being \ha, because we have already detected \oii\
emission with our narrowband imaging. The FMOS spectra demonstrate
that our selection of NB912 and NB921 \oii\ emitters is valid and that
the structure we have found around the cluster is real.     

We also have additional spectroscopic redshifts from the literature. 
\citet{hilton2010} have spectroscopically confirmed 44 member galaxies
in this cluster. We cross-identify the galaxies confirmed by
our Subaru observations to the 44 member galaxies in
\citet{hilton2010}, and seven galaxies turn out to be common 
objects. We checked that we do not find any disagreement in the
redshifts of the common sources. Therefore, 101 galaxies are
spectroscopically confirmed to be associated with the structure in and
around the galaxy cluster. The redshift distribution of the confirmed
galaxies is shown in Figure \ref{fig;NBfilters}.    

\section{Selection of emission line galaxies}
\label{sec;catalogues}

\begin{figure*}
 \begin{center}
 \includegraphics[width=0.9\linewidth]{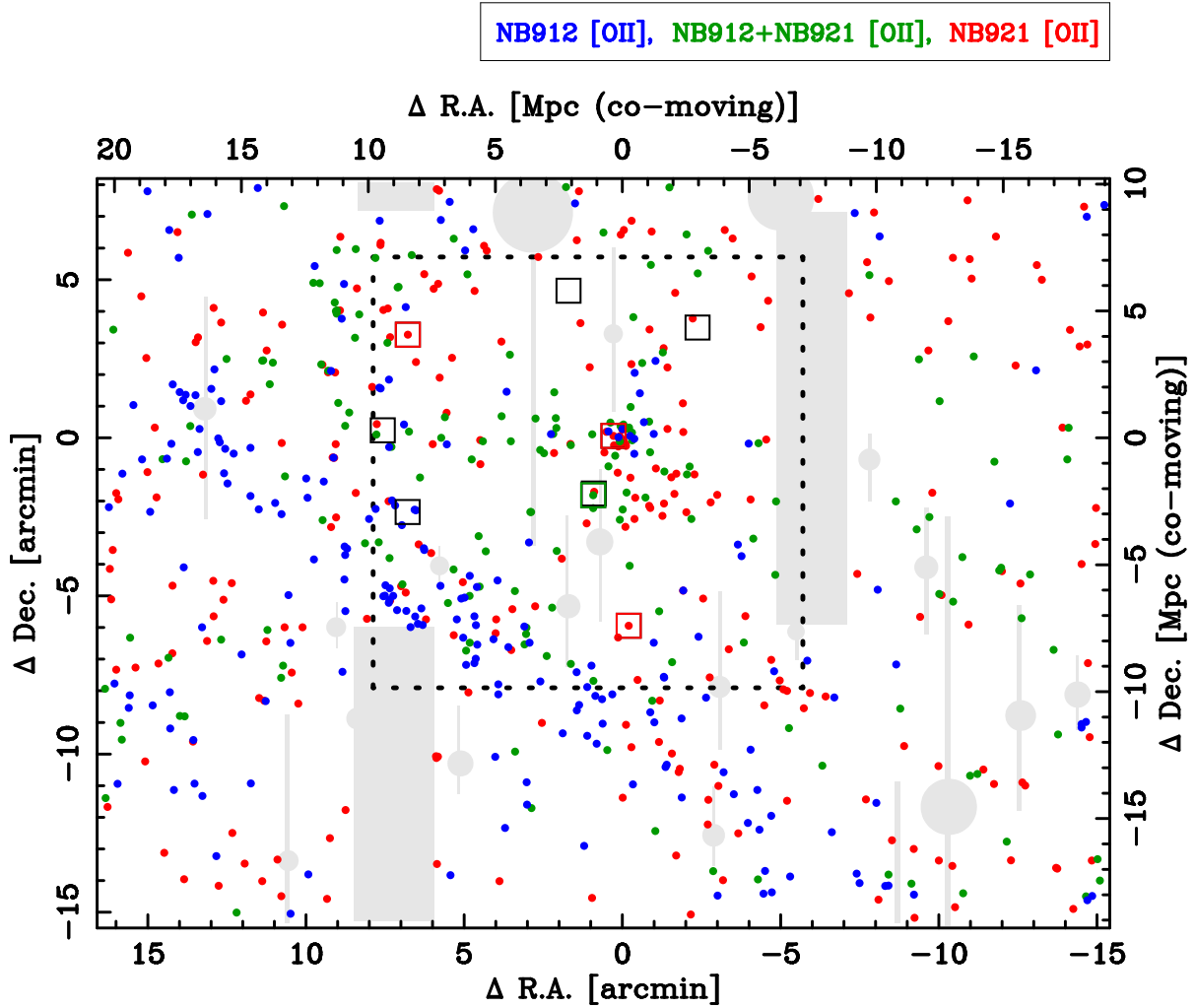}
 \end{center}
 \caption{ Celestial distribution of NB912/NB921 \oii\ emitters and
   \NBH\ \ha\ emitters. Blue dots show \oii\ emitters identified from
   NB912 data (i.e., NB912 \oii), while red dots show those from NB921 
   data (i.e., NB921 \oii). Green dots are \oii\ emitters identified
   from both NB912 and NB921 data (i.e., NB912+NB921 \oii). 
   Solid open squares show \NBH\ \ha\ emitters, following the same
   colour-coding as for \oii\ emitters (e.g., the \ha\ emitters
   identified as NB921 \oii\ emitters as well are shown by red open squares). 
   Black open squares indicate \ha\ emitters without any \oii\ emission
   detected.  
   The gray regions are masked due to the bad quality of the images,
   and the dotted square shows the FoV of a single detector of WFCAM on
   UKIRT (see also the text in \S~\ref{sec;HiZELSdata}). 
   The axes shows relative coordinates from the centre of the galaxy
   cluster. As already found in \citet{hayashi2011}, we can confirm
   again the clustering of \oii\ emitters in the centre and the large
   scale structure existing in the surrounding region even when the
   NB921 data are used.   
  }   
 \label{fig;map}
\end{figure*}

In this section, we select emission line galaxies using the
photometric catalogues. We have already published the sample of \oii\ 
emitters at $z\sim1.46$ based on the NB912 data \citep{hayashi2011}. 
Since there is only a slight difference between NB912 and NB921 in
the central wavelength of response curve, we can apply the same method
for the selection of NB921 \oii\ emitters as for NB912 \oii\ emitters
summarised below. We refer readers to \citet{hayashi2011} for more
details, if needed.   

First, we select galaxies with colour excess in $z'$-NB921 at more
than 3$\sigma$ (Figure~\ref{fig;selection}). This corresponds to
galaxies with a line flux larger than 1.2$\times$10$^{-17}$ erg
s$^{-1}$ cm$^{-2}$ being selected. If the excess is due to the \oii\
emission line at $z=1.46$, the limiting flux corresponds to a
dust-free SFR of 2.2 \Msun\ yr$^{-1}$ according to the \oii--SFR
calibration in \citet{kennicutt1998} where a \citet{Salpeter1955}
initial mass function (IMF) is assumed.    
To exclude the possible contamination of galaxies due mainly to
photometric errors, additional criterion for the observed equivalent
width (EW) larger than 35\AA\ (i.e., 14\AA\ in the rest-frame for \oii\ at
$z=1.46$) is applied. As a result, we select 1,135 NB921
emitters. Among them, 789 emitters are detected in $K$ at more than
2$\sigma$ level. In what follows, we deal with the $K$-detected
emitters, because we identify the \oii\ emitters based on their $B-z'$
and $z'-K$ colours \citep{hayashi2011}. 
With the criteria, we select 429 NB921 \oii\ emitters in total
(Figure~\ref{fig;selection}). We note that the 
number of NB921 \oii\ emitters is slightly larger than that of NB912
\oii\ emitters even if the same limiting \oii\ flux is adopted between
the two samples. However, the numbers of \oii\ emitters selected are
in agreement within a 1 $\sigma$ error based on Poisson statistics. 
The small difference can appear because of the structures and
clustering of galaxies at slightly different redshifts.

We can use the \NBH\ and H images to select \ha\ emission line
galaxies at $z\sim1.46$. Again, galaxies with a colour excess in
$H$--\NBH\ are selected, and then the same colour selection ($Bz'K$)
is applied to identify \ha\ emitters among the 19 emission line candidates. 
The cut of 3$\sigma$ colour excess corresponds to galaxies with a line
flux larger than 1.4$\times$10$^{-16}$ erg s$^{-1}$ cm$^{-2}$ being
selected. If the excess is due to \ha\ at $z=1.46$, the limiting flux
corresponds to a dust-free SFR of 14.6 \Msun\ yr$^{-1}$ according 
to \citet{kennicutt1998}. We also apply an observed EW cut larger than
50\AA\ (i.e., 20\AA\ in the rest-frame for \ha\ at $z=1.46$). 
Although \citet{Sobral2013} made a small correction to $H$--\NBH\
colours using $J-H$ colours to estimate the continuum level in the
broadband more accurately, we find that $H$--\NBH\ colours are
distributed around zero without any correction.
Thus, no further correction is applied for the colour term and the
method used for the selection is similar to that applied for the
selection of \oii\ emitters at $z\sim1.46$. With their criteria, we
select 9 \ha\ emitters in total (Figure~\ref{fig;selection}).       

Now, we have three samples of emission line galaxies at $z\sim1.46$;
NB912 \oii\ emitters, NB921 \oii\ emitters, and \NBH\ \ha\ emitters.
Since they are all at similar redshifts, some emitters in the
catalogues should overlap each other. First, we compare the catalogues
between NB912 and NB921 \oii\ emitters by searching for galaxies for
which the difference in the coordinates are within 0.5 arcsec. 
Among 380 NB912 and 429 NB921 \oii\ emitters, it is found that 170
galaxies are common in both samples. We also find that four \NBH\ \ha\
emitters are in the \oii\ emitter samples by using a circle with 1.0
arcsec radius to search for the common galaxies. Moreover,
cross-matching with the spectroscopic data described in
\S~\ref{sec;specdata}, we spectroscopically confirm 41 NB912+NB921, 21
NB912, 11 NB921 \oii\ emitters and two \ha\ emitters
(see also Figure \ref{fig;NBfilters}).    

Figure~\ref{fig;map} shows a celestial distribution of NB912 or NB921
\oii\ emitters. It is worth mentioning that we can confirm
again the clustering of \oii\ emitters in the centre and the
large-scale structure existing in the surrounding region which were
already found in \citet{hayashi2011}, even when the NB921 data are used. 
More importantly, it seems that there is a slight difference in the
distribution between NB912 and NB921 \oii\ emitters,
although about half of them overlap.
The prominent filamentary structure that \citet{hayashi2011} have
discovered is likely to be slightly in the foreground, and the
distribution of NB921 \oii\ emitters seems to be spread along the
southwestern direction from the cluster centre. 
As shown in Figure~\ref{fig;NBfilters}, the NB912 and NB921 can detect
galaxies with a difference in velocity by $\sim$2000 km s$^{-1}$.
Thus, all of the differences suggest that there is a three dimensional
structure around the galaxy cluster. 
A similar result has been reported at higher redshift as well. 
\citet{shimasaku2004} used the samples of Lyman $\alpha$ Emitters
(LAEs) selected by two narrowband filters and found that the
distribution of the galaxies are quite different at $z=4.79$ and 4.86.    
These observations demonstrate that revealing the three dimensional
structure is crucial to improve our understanding, in particular in
the early Universe when galaxies are active and the structures are
growing. This implies the two dimensional celestial distribution could
lead to errors in estimating the environment in which galaxies reside 
due to projection effects. This is investigated in the following
sections using the two \oii\ emitter samples.   

Figure~\ref{fig;map} also shows a celestial distribution of \NBH\ \ha\
emitters. Unlike the \oii\ emitters, \ha\ emitters are not clustered,
but this is a likely consequence of the significantly shallow
detection limit of \ha\ emission when compared to \oii. However, it
seems that the distribution of \ha\ emitters is consistent with
tracing that of \oii\ emitters.

\section{Redshift measurements with two adjacent narrow-band filters}
\label{sec;redshifts}

\begin{figure}
 \begin{center}
 \includegraphics[width=\linewidth]{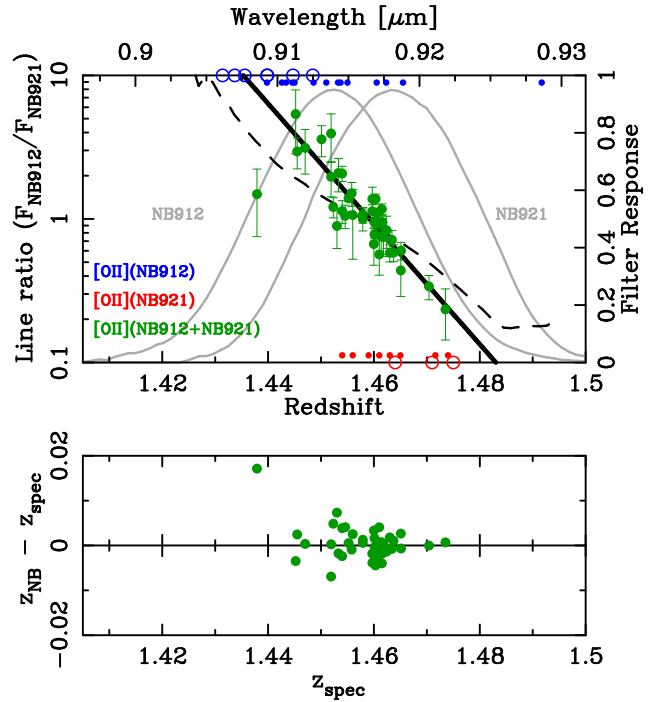}
 \end{center}
 \caption{ Top: Ratio of \oii\ emission lines measured with
   NB912 filter to that measured with NB921 filter as a function of
   redshift of galaxy. The broken line shows the line ratio expected
   from the difference of the response function between the narrowband 
   filters. Filled green circles show the measured line ratios for
   the NB912+NB921 \oii\ emitters with spectroscopic redshift. 
   The solid line shows the linear relation between the line ratio and
   redshift obtained from the fitting to the data of the NB912+NB921
   \oii\ emitters. 
   Blue(red) symbols show NB912(NB921) \oii\ emitters without the
   detection of \oii\ emission in NB921(NB921), respectively. 
   Among them, open circles shows the galaxies having \oii\ flux
   inadequate to be detected in the other narrowband, since the
   expected flux is smaller than the detection limit.
   In addition, the response functions of the two filters are shown by
   gray curves. 
   Bottom: The comparison between the redshift, $z_{\rm NB}$, estimated from
   the line flux ratio measured by NB912 and NB921 filters and the
   spectroscopic redshift, $z_{\rm spec}$, for the NB912+NB921 \oii\
   emitters as a function of spectroscopic redshift. 
   See the text (\S~\ref{sec;redshifts}) for details of how the
   redshift, $z_{\rm NB}$, is estimated using the two narrowbands. 
   The standard deviation of the difference is $\sigma((z_{\rm NB}-z_{\rm spec})/(1+z_{\rm spec}))=0.002$.
  }   
 \label{fig;fluxratio}
\end{figure}

\begin{figure}
 \begin{center}
 \includegraphics[width=\linewidth]{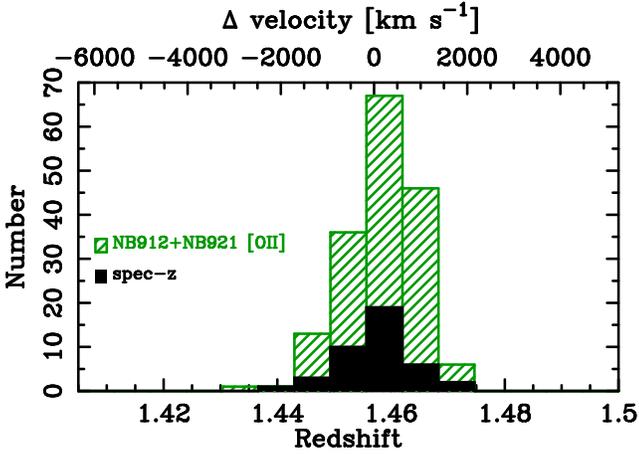}
 \end{center}
 \caption{ 
   The distribution of the redshift estimated from the line ratio
   measured by NB912 and NB921 filters for the NB912+NB921 \oii\
   emitters, is shown by the hatched histogram. Among them, \oii\
   emitters with the spectroscopic redshift are shown by the filled
   histogram. The two histograms are not considered to be clearly
   different distributions, based on the K-S test, suggesting
   that the method we use to estimate the redshift is valid.
  }   
 \label{fig;Nzoii}
\end{figure}

The two narrowband filters of NB912 and NB921 on Suprime-Cam are
very close to each other (Figure~\ref{fig;NBfilters}). The overlap with a
slight difference in the response curves allows us to estimate the
redshift where the \oii\ emitters are located between
$z\approx$1.42--1.49, based on the difference of  
emission line fluxes measured in the NB912 and NB921 images. This is
because the narrowband filters are not perfect top-hats.
That is, the flux measured by narrowband imaging for a galaxy
with a given luminosity is dependent on its wavelength. For example,
if a star-forming galaxy is at $z=$1.44, it should be detected as a
NB912 emitter but not detected as a NB921 emitter. On the other hand,
if a star-forming galaxy is at $z=$1.48, it should be detected as a
NB921 emitter but not detected as a NB912 emitter. More interesting
are star-forming galaxies detected with both NB912 and NB921
images. If the ratio of line flux measured with NB912 to that with
NB921 is greater than unity, the galaxy is expected to be at $z<1.46$,
while if the ratio is smaller than unity the galaxy is likely to be at
$z>1.46$. Therefore, we can reveal the three-dimensional large scale
structure in and around the XCS2215 galaxy cluster with these two
filters.   

Figure \ref{fig;fluxratio} shows that the expected ratio of
NB912 to NB921-measured line flux is a monotonically decreasing
function of redshift between $z=1.430$ and $z=1.485$ (shown by the
broken line), 
demonstrating that we can estimate the redshift from the flux ratio
measured with the two narrowband filters. We then make sure of the
validity of the method using the \oii\ emitters which have been
spectroscopically confirmed. For NB912+NB921 \oii\ emitters, it is
clearly seen that the flux ratio is correlated with the spectroscopic
redshift and the flux ratios measured are distributed around the
values expected from the response function of the narrowband
filters. However, a systematic difference between the actual line
ratio and the expectation is seen when the redshift is far from
$z\sim1.46$. The discrepancy can be caused by inhomogeneity of
response curve over the FoV and/or the difference between the actual
response curve when installed on the Suprime-Cam and the measurement
in a laboratory, although it is unlikely that the response curve of
the two filters have large inhomogeneity. 
To correct for those effects, we fit a linear
function to all of the data of NB912+NB921 \oii\ emitters, and the
fitted relation is shown by the solid line in Figure \ref{fig;fluxratio}. 
Hereafter, we use this linear relation obtained from the fitting to estimate
the redshift of \oii\ emitters. Note that we removed the
NB912+NB921 \oii\ emitter at $z=1.438$ from the fitting since the deviation from
the others is large, but even if the emitter is included, we make sure
that the results shown below are not largely changed. 

The \oii\ emitters identified in only NB912 tend to have spectroscopic
redshifts less than $z\sim1.46$ which corresponds to the 
redshift where the galaxy has the NB912/NB921 line ratio of unity,
while those identified in only NB921 tend to have the redshift greater
than $z\sim1.46$. This is also what we expect. Indeed, the
blue(red) open circles show the NB912(NB921) \oii\ emitters for which
the expected NB921(NB912) \oii\ flux is smaller than the detection
limit. This is the case for about a third of NB912 or NB921 \oii\
emitters. Another third of \oii\ emitters seem to have \oii\
fluxes which in theory should be able to be detected in both
narrowbands. However, the expected \oii\ fluxes in the non-detection
narrowband are close to the detection limit. Thus, the actual \oii\
flux may be smaller than the limit due to the uncertainty of \oii\
flux measured by the other narrowband. Finally, the remaining third of
\oii\ emitters should have a \oii\ flux which is easily detected in
both narrowbands, and yet are absent in one narrowband. Although
the cause is unclear, the criteria of EW and significance of colour
excess applied to select the emitters as well as the colour selection
to identify the emission line can influence the detection of \oii\
emission. Nevertheless, the redshift can be estimated from the line
flux for about 70\% of \oii\ emitters detected by NB912 or NB921
narrowband filters, which adequately allows us to trace the three
dimensional structure.

The redshift estimated from the line ratio measured by NB912 and NB921
filters, $z_{\rm NB}$, is compared with the spectroscopic redshift,
$z_{\rm spec}$, for the NB912+NB921 \oii\ emitters. We then calculate
the difference of the two redshifts and find that 
$\sigma((z_{\rm NB}-z_{\rm spec})/(1+z_{\rm spec}))=0.002$ (Figure~\ref{fig;fluxratio}). 
This suggests that the accuracy of the redshift, $z_{\rm NB}$, is
satisfactory, and despite only two narrowband datapoints being used,
it is better than that of the photometric redshifts estimated from
photometry with as many as 30-bands in the COSMOS field;
$\sigma_{\Delta z/(1+z_{\rm s})}=0.007$ at $i^{+}_{\rm AB}<22.5$, 
$\sigma_{\Delta z/(1+z_{\rm s})}=0.012$ at $i^{+}_{\rm AB}<24$ and $z<1.25$,
and $\sigma_{\Delta z/(1+z_{\rm s})}=0.06$ at $i^{+}_{\rm AB}\sim24$
and $z\sim2$ \citep{ilbert2009}. 
Thus, the line ratio measured with the adjacent
narrowband filters can be a powerful estimator of redshift enabling us
to map out well the three dimensional structure of star forming
galaxies in the field surveyed without spectroscopy of {\it all}
galaxies.    

Figure~\ref{fig;Nzoii} shows the distributions of the two redshift
estimations; $z_{\rm NB}$ and $z_{\rm spec}$. The histograms seem to be 
similar to each other. Indeed, the Kolmogorov-Smirnov (K-S) test
indicates that they are likely to show the same distribution and 
the probability of their being based on the same populations is more than 70\%. 
Thus, we conclude that we can use the line ratio to estimate the
redshift of \oii\ emitters and then investigate their three
dimensional distribution in and around the galaxy cluster quite
accurately.

\section{Results}
\label{sec;results}

\subsection{3-D structures around the XCS2215 cluster}
\label{sec;3ds}

\begin{figure*}
 \vspace{-0.5cm}
 \begin{center}
   \includegraphics[width=1.2\linewidth]{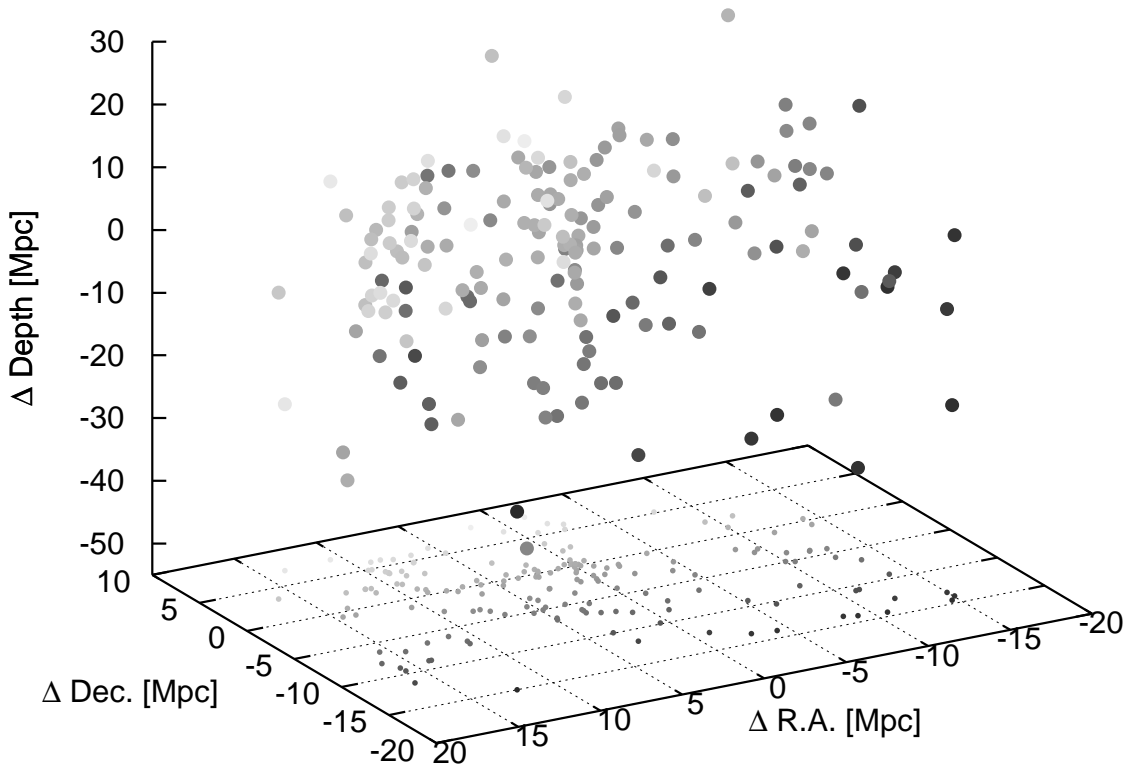}
 \end{center}
 \vspace{-1.5cm}
 \caption{ 
   Three dimensional distribution of NB912+NB921 \oii\
   emitters. The depth in z-axis is converted from the redshift, which
   is estimated from the ratio of \oii\ emission lines measured with
   NB912 filter to that measured with NB921 filter
   (Figure~\ref{fig;fluxratio}). 
   The filled circles are shown in gray scale based on the declination;
   darker colours mean lower declination.    
   The filamentary structures are seen in three dimensional
   space. The dots in the R.A.-Dec.~plane are the projection on the
   celestial plane where the coordinates are shown in co-moving scale
   relatively to the centre of the galaxy cluster. The radius of
   $R_{200}$ in the XCS2215 cluster is 2.0 Mpc in co-moving scale \citep{hilton2010}.
  }   
 \label{fig;3d}
\end{figure*}

\begin{figure*}
  \vspace{-1.0cm}
  \hspace{-1.8cm}
  \begin{tabular}{cc}
    \begin{minipage}{0.5\hsize}
      \begin{center}
        \includegraphics[width=1.3\linewidth]{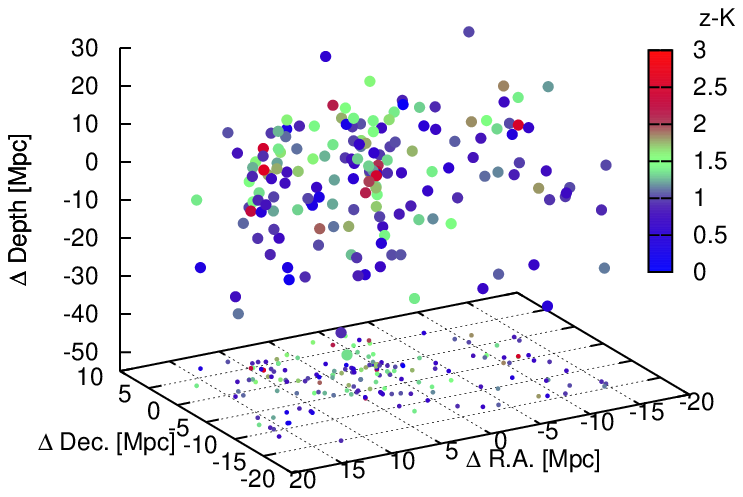}
      \end{center}
    \end{minipage}
    \begin{minipage}{0.5\hsize}
      \begin{center}
        \includegraphics[width=1.3\linewidth]{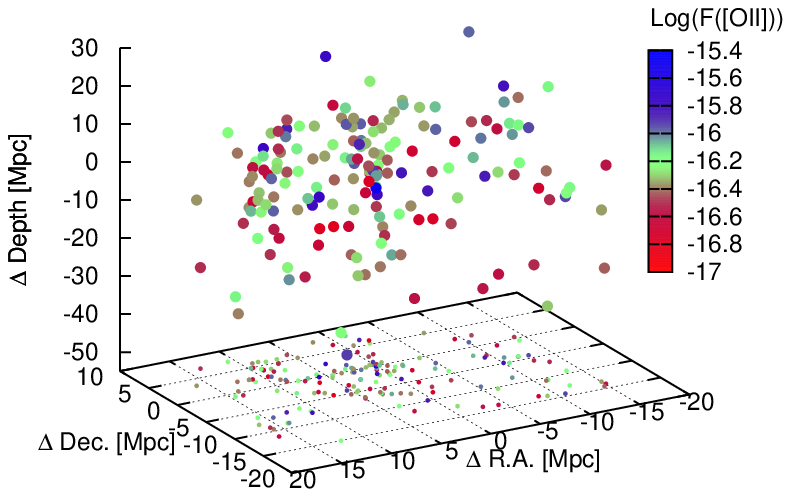}
      \end{center}
    \end{minipage}
  \end{tabular}
    \vspace{-0.5cm}
  \caption{ Same as Figure~\ref{fig;3d}, but the symbols are
   colour-coded by $z-K$ colour in the left-hand panel and \oii\ flux
   in the right-hand panel. The $z-K$ colour of galaxies at $z\sim1.46$
   corresponds to the $U-z$ colour in rest-frame, implying that it is
   useful to estimate the strength of the Balmer/4000\AA\ break. The \oii\
   fluxes are corrected for the response curve of the filter using the
   redshift estimated for the ratio of NB912 to NB921 flux, and the
   luminosity of the \oii\ emission line is an indicator of the star
   formation rate of the galaxy.  
  }   
 \label{fig;3d_zkoii}
\end{figure*}

Figure~\ref{fig;3d} shows the three dimensional distribution of
NB912+NB921 \oii\ emitters. Although we have seen the filamentary
structures in this field (Figure~\ref{fig;map} and
\citet{hayashi2011}), the 3D view reveals that it is not an effect of
projection but a real distribution and there are several new
filamentary structures. It seems that some structures extend
towards the centre of the galaxy cluster, and some intersect
other structures. These structures cannot be fully understood without
seeing the 3D distribution of galaxies as well as the projected
celestial distribution.  
Galaxy clusters are considered to be formed at the intersection of
filamentary structures and grow while assembling galaxies by the
gravity of the system along the large-scale structures. The 3D
structures we have found around the galaxy cluster is fully consistent
with this picture. We should be witnessing the site where galaxies in
the surrounding region are infalling towards the centre of galaxy
cluster at $z\sim1.5$ when the vigorous evolution of galaxies occurred. 

Since we use the redshift to draw the distribution of star-forming
galaxies in three dimension, we should take care of the effect of
peculiar motion which distorts that view. In
particular, this effect could work significantly very near the cluster
core and the structure in the redshift direction (i.e., depth) could
seem to shrink relative to what it is. However, there is no doubt of
the existence of the large scale structure, since the filamentary
structures tend to spread widely in the direction parallel to the
celestial plane.  

Figure~\ref{fig;3d} resembles the structures seen
in cosmological $N$-body simulations based on the CDM model
\citep[e.g.,][]{Springel2005, ishiyama2013}. The simulations predict the 
distribution of dark matter. However, the consistency between the
cosmological simulation and the observations at $z\sim1.5$ confirms
that the distribution of galaxies in the Universe is governed by dark
matter and grows hierarchically. 
In addition, the three dimensional structure of the galaxy cluster at
$z=1.46$ provides us with an unique opportunity to investigate the
dependence of galaxy properties on the location of galaxies infalling
into the cluster.  
At lower redshifts, similar large-scale structures have been
found \citep[e.g.,][]{kodama2005,Nakata2005,tanaka2009,Sobral2011,gal2008,Faloon2013}. 
However, the properties of galaxies associated with the cluster is
different, showing that star formation activity weakens in the core. It
is interesting to investigate what causes the change of the properties
in the course of galaxies infalling along the filamentary
structure. We discuss this point in \S~\ref{sec;env_dependent_properties}.  

\subsubsection{Comparison between \oii\ and \ha\ emitters}

Thanks to the unique set of narrowband filters, we can compare \ha\
emission with \oii\ emission in an unbiased manner for galaxies
associated with the structures at $z\sim1.46$. Similar comparisons
have been conducted for emission line galaxies detected with dual
narrowband filters in the field \citep{Sobral2012,hayashi2013}. Before
comparing them, we must keep in mind that the limiting fluxes in \ha\
and \oii\ emissions are quite different as shown in
\S\ref{sec;catalogues} or Table \ref{tbl;NBfilters};
1.4$\times$10$^{-16}$ erg s$^{-1}$ cm$^{-2}$ for \ha\ emitters (a
dust-free SFR of 14.6 \Msun\ yr$^{-1}$) and 1.2 or
1.4$\times$10$^{-17}$ erg s$^{-1}$ cm$^{-2}$ for \oii\ emitters (a
dust-free SFR of 2.2 or 2.6 \Msun\ yr$^{-1}$). Also, note that this
\ha\ limit is much shallower than those in \citet{Sobral2012} and/or
\citet{hayashi2013}. 

As shown in Figure~\ref{fig;map}, for four out of nine \ha\ emitters,
\oii\ emissions are detected by NB912 or NB921 imaging. All of them
have \oii\ emission detected in NB921 data, but only one emitter has
\oii\ emission detected in NB912. This is likely due to the difference
between the filter response curves in redshift space (Figure~\ref{fig;NBfilters}). 
Note that among the four, the \ha\ emitter with \oii\ emission
detected in both NB912 and NB921 is spectroscopically confirmed, and
the others are not confirmed yet. 
Figure~\ref{fig;selection} shows
that the \ha\ emitters with \oii\ detections tend to have redder
$z'-K$ colour. This may go against our intuition. We expect that
\oii\ emission is easier to detect in less dusty galaxies which
should have bluer colour. The observations seem to suggest that the
dust extinction does not determine mainly whether the \oii\ emission
is detected or not for the \ha\ emitters.   
Perhaps, it is related to the age, if the redder $z'-K$ colour means
stronger Balmer/4000\AA\ break due to an older stellar population
rather than heavier dust extinction. 
We check the EW of the emitters, but we do not find any significant
difference between the \oii\ emitters with and without \ha\ detection
(see also Figure \ref{fig;selection}). 
The ratio of \ha/\oii\ is also dependent on metallicity
\citep[e.g.,][]{Sobral2012,hayashi2013}. 
If the galaxies have lower metallicity, \oii\ luminosity should be higher
and easier to be detected for galaxies with a given \ha\ luminosity.
On the other hand, about half of \ha\ emitters do not have \oii\
emissions detected in NB912 or NB921 data. These galaxies may be
metal-rich.
The galaxies detected as a \NBH\ emitter could alternatively be \oii\
emitters at $z=3.3$ and/or \oiii\ emitters at $z=2.2$. However, since
the fluxes of the emission lines detected are greater than
1.4$\times$10$^{-16}$ erg s$^{-1}$ cm$^{-2}$, the corresponding
luminosities are quite large if the galaxies are at $z\gg2$. Thus, it
is unlikely that almost all of them are contaminants.   

In comparison with the large numbers of \oii\ emitters being detected,
the number of \ha\ emitters detected is small. Typically, \oii\
emitters at $z\sim1.46$ are a population with a lower amount of dust
\citep{hayashi2013}. Less dusty galaxies have an observed ratio of
\ha\ to \oii\ of $\sim1$, suggesting that \ha\ luminosity may not be
strong for galaxies with a given \oii\ luminosity. Indeed, there are
only four \oii\ emitters with \oii\ fluxes larger than
1.4$\times$10$^{-16}$ erg s$^{-1}$ cm$^{-2}$ in the region where \NBH\
and $H$ data by WFCAM are available. Even if the limit is weakened
down to 1.0$\times$10$^{-16}$ erg s$^{-1}$ cm$^{-2}$, the number of
the \oii\ emitters is only 12. This is of the same order of magnitude
as the number of the \ha\ emitters detected among the \oii\ emitter
sample. This means that the ratio of \ha\ to \oii\ 
is not high but close to unity, which may support that most of the
\oii\ emitters are typically less dusty. Thus, a deeper survey of \ha\
emission is required so that we can obtain a larger sample of \ha\
emitters at $z\sim1.5$.

\subsection{Environmental dependence of galaxy properties}
\label{sec;env_dependent_properties}

One of the hot topics in the evolution of cluster galaxies is how and where
the galaxies obtained the properties characteristic of cluster galaxies 
seen in the local Universe, with redder colour, less star formation
activity, and bulge-dominated morphology 
\citep[e.g.,][]{Blakeslee2006,Tanaka2005,Tanaka2013,Gerke2007,vandenBosch2008,Thomas2010,Lemaux2012,RodriguezDelPino2013,Newman2013,Strazzullo2013,Bassett2013}. 
Some effects such as ram pressure stripping or galaxy-galaxy
interactions are expected to be at work for infalling galaxies in the
outskirts \citep[e.g.,][and references therein]{Treu2003,Moran2007}. 
It is therefore interesting to investigate how the
properties of \oii\ emitters change along the filamentary large scale
structures. 

Figure~\ref{fig;3d_zkoii} shows again the three dimensional
distribution of the 170 \oii\ emitters identified by both narrowband 
filters, but now colour-coded by $z-K$ colour and \oii\ flux. 
Note that the $z-K$ colour of galaxies at $z\sim1.46$ corresponds to
the $U-z$ colour in rest-frame, implying that it is useful to estimate the
strength of the Balmer/4000\AA\ break.
It seems that \oii\ emitters with red or green $z-K$ colour of
$\ga1.3$ are likely to be located along the filamentary structures and
in the region close to the cluster centre,
where the criterion of red or green colour is based on the
distribution of $z-K$ colours shown in Figure~\ref{fig;selection}. 
The \oii\ emitters away from the cluster centre, and relatively
isolated ones, seem to have bluer colours. If this is true, it can be
considered that the galaxies have changed colour 
while they are in the filamentary structures and close to the centre
of galaxy cluster.  
On the other hand, no strong dependence of \oii\ flux on the location
of the galaxy is seen, although we might have expected that the star
formation of galaxies weakens along the structures.  
We note that stellar mass of the galaxies also does not show a strong
dependence. This indicates that there is no strong dependence of
specific SFRs on the location of the galaxy, which is consistent with
the result that \citet{Brodwin2013} have found for galaxy clusters at
$z>1.4$. Although the result may be somewhat subjective, we discuss
the dependence of galaxy properties on the environment more
quantitatively below. 

The red colour of galaxies can be caused by two degenerate
factors. One is the age of the galaxy and the other is dust. If we
assume that the age mainly dominates the change of the colour, since
\oii\ emitters at $z\sim1.46$ are likely to be typically less dusty
populations \citep{hayashi2013}, then the galaxies get older 
as they are in the filamentary structures and close to the centre of
galaxy cluster.  
If we assume that dust extinction is a dominant cause of
their red colors, this means that dusty starburst activity is
preferentially triggered 
in the environments of filamentary structures and the centre of the
galaxy cluster.    
This is very similar to our recent finding by
\citet{koyama2013b}, where we demonstrated that star-forming galaxies
in dense environments tend to be more highly obscured by dust than
normal field galaxies at the same redshifts. 
Thus, in any case, our result suggests that galaxy evolution occurs
along the large-scale filamentary structures around the galaxy
cluster. Furthermore, if we assume that galaxies in the surrounding
region follow the scenario of hierarchical cluster formation as
expected, galaxies are considered to evolve while infalling to the
centre along the filamentary structures.    

A dependence of the star formation activity on the site where galaxies
reside is not seen in this cluster. However, studies of galaxy
clusters at lower redshifts suggest that galaxies in the core of rich
clusters must stop their activity significantly before $z\sim1$. Thus,
quenching process might be at work in the core region more
effectively. \citet{hayashi2011} have found an excess of \oii\ 
emitters with red colours comparable to galaxies on the red sequence
of colour-magnitude diagram, and they are likely to be possible
AGNs. It is argued that the AGN feedback could play a role in
quenching star formation of galaxies in the core of the galaxy cluster.   

\begin{figure}
 \begin{center}
 \includegraphics[width=\linewidth]{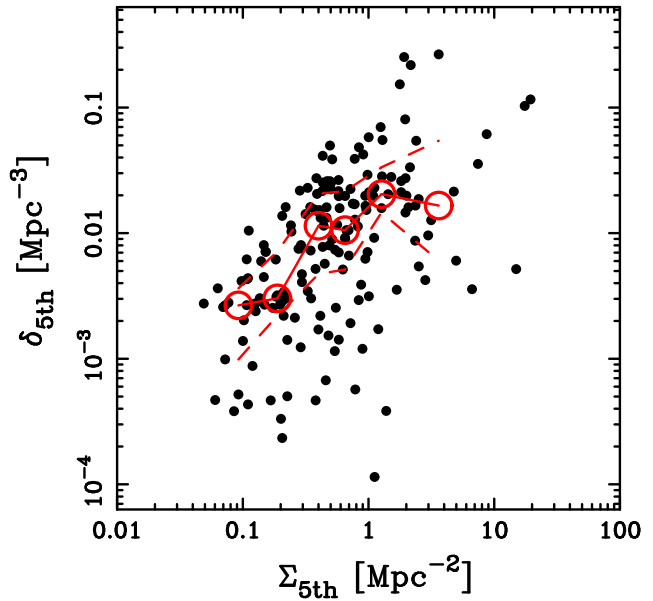}

 \end{center}
 \caption{The comparison of the local density, $\delta_{\rm
     5th}$, calculated from the three dimensional distribution with
   the surface local density, $\Sigma_{\rm 5th}$, in
   \citet{hayashi2011} for the NB912+NB921 \oii\ emitters. 
   The solid line and open circles shows the median values in
   each bin, and the broken lines show the 25\% and 75\% values.   
  }   
 \label{fig;comp_density}
\end{figure}

\begin{figure*}
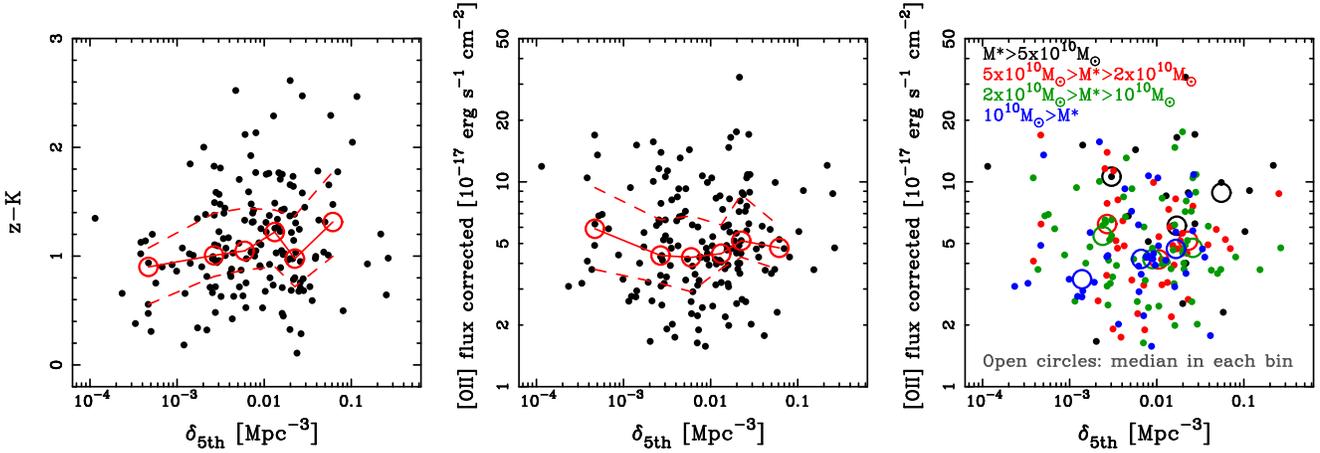

 \begin{center}
 \includegraphics[width=0.33\linewidth]{./fig9a.eps}
 \includegraphics[width=0.33\linewidth]{./fig9b.eps}
 \includegraphics[width=0.33\linewidth]{./fig9c.eps}

 \end{center}
 \caption{ (Left) The $z-K$ colours of the galaxies as a function of
   the local density, $\delta_{\rm 5th}$, for the NB912+NB921 \oii\
   emitters. The solid line and open circles show the median values
   in each bin, and the broken lines show the 25\% and 75\% values.    
   (Middle) The same as the left panel, but for the \oii\ fluxes
   corrected for the response curve of the filter using the redshift
   estimated for the ratio of NB912 to NB921 flux. 
   (Right) The same as the middle panel, but the symbols are
   colour-coded based on the stellar mass; black for galaxies with M$_{\rm stellar}$
   $>5\times10^{10}$\Msun, red for galaxies with M$_{\rm
     stellar}=$2--5$\times10^{10}$\Msun, green for galaxies with M$_{\rm
     stellar}=$1--2$\times10^{10}$\Msun, blue for galaxies with M$_{\rm
     stellar}<10^{10}$\Msun.
   Open circles show the median of \oii\ fluxes in each bin of the
   local density. 
  }   
 \label{fig;env_properties}
\end{figure*}

Here, we calculate a local density from the three
dimensional distribution (Figure~\ref{fig;3d}) to discuss the
dependence of galaxy properties on the environment more
quantitatively. We note that similar discussions were made in
\citet{hayashi2011}, but we can refine them with the three dimensional
distribution of the galaxies.     
The local density, $\delta_{\rm 5th}$, in units of Mpc$^{-3}$ (in
co-moving scale), is calculated using the volume out to which the
fifth nearest \oii\ emitters are included. We then compare this with
the local surface density, $\Sigma_{\rm 5th}$, in units of Mpc$^{-2}$ 
(in co-moving scale), calculated in \citet{hayashi2011}. 
Note that the NB912+NB921 \oii\ emitters are used in the calculation of 
$\delta_{\rm 5th}$, while all of the NB912 \oii\ emitters are used in
the calculation of $\Sigma_{\rm 5th}$ in \citet{hayashi2011}. However,
as shown in Figure \ref{fig;fluxratio}, many of the NB912 \oii\
emitters without a \oii\ detection in NB921 have redshifts lower than
$z\sim1.45$, that is, they tend to be embedded in structures slightly
in the foreground compared with the galaxies associated with the
cluster. Thus, although the samples used are not identical, the
comparison allows us to investigate the properties of galaxies in the
cluster and the structures at $z\sim1.46$. 
We also make sure that our results do not change even if the local
densities are calculated from the third or seventh nearest
emitters. In this work, we adopt the local densities calculated from
the fifth nearest emitters to be consistent with \citet{hayashi2011}.  
Figure~\ref{fig;comp_density} shows the comparison between the two  
local densities. 
Although the two densities are roughly in agreement, the correlation
is not strong and there is a large dispersion between them. Indeed,
the correlation coefficient is 0.52. The mild correlation between
them underlines the importance of obtaining the true three dimensional
distribution to properly estimate the environment where galaxies
reside, instead of using a projected celestial distribution.    
However, as described in \S~\ref{sec;3ds}, the peculiar motion of
an individual galaxy can influence the local density estimated from
the three dimensional distribution whose depth is based on the
redshift.   

Figure~\ref{fig;env_properties} shows the $z-K$ colours and \oii\
fluxes as a function of local density, 
$\delta_{\rm 5th}$, where the fluxes are corrected for the response
curve of the filter using the redshift estimated from the ratio of
NB912 to NB921 flux. The correction to the flux varies by a
factor of 1.00 to 8.83, and the median is a factor of 1.06. Note that
the precise correction is dependent on the redshift at which the
galaxies are located (Figure~\ref{fig;NBfilters}). Our method using
two adjacent narrowband filters demonstrates that we can measure both
accurate fluxes of the emission lines and accurate redshifts. In both
panels, the median, 25\% and 75\% values are also shown in each
bin. The distribution of the colours and \oii\ 
fluxes in each panel is consistent with our earlier discussion.
First, the average colour of galaxies mildly 
gets redder as the local density becomes larger, although the 
difference in colour is not large. Moreover, almost all galaxies with
red colours of $z-K>2.2$, as defined in \citet{hayashi2011}, tend to be
located in higher density regions, while no such red galaxy is seen in
the lower density regions. 
Next, there is no strong dependence of \oii\ flux on the local
density, suggesting that the SFRs of galaxies associated with this cluster
and structures are not dependent on the environment. Note that the
\oii\ fluxes are not corrected for dust extinction. However, even
if a dust correction dependent on SFR or stellar mass is adopted for the
\oii\ emitters, the result does not change, as we do not observe a
strong dependence of stellar masses of galaxies on the local density.  
We also investigate the star-formation activity in galaxies at a given
stellar mass as a function of the local density. 
There is no strong dependence of star formation on environment in all
bins of stellar mass. Consequently, although we find that the surface
local density is not an ideal indicator of the environment, the
results we have found in this paper support those of \citet{hayashi2011}. 

In summary, our results show that there is a weak correlation between
the galaxy colour and the environment on average, despite other
properties showing no environmental dependence. As discussed above,
galaxy colours can relate to the age or dust extinction. In the
former case, it is likely that galaxies have experienced a similar,
steady history of star formation in all environments, but galaxies
formed earlier on average in denser environments. We may be witnessing
the formation bias of cluster galaxies.    
In the latter case, it is also likely that star formation activity is
underestimated and the dusty starburst might have something to
do with the mode of star formation which is triggered by the
environmental effects and thus different from that of field
galaxies. It would be interesting to reveal which is the main factor
to cause the environmental dependence of galaxy properties we have
observed. However, such investigation is beyond the scope of this
paper due to the limitations of the data.    

\section{Conclusions}
\label{sec;conclusions}

We have revealed large-scale structures around a galaxy cluster,
XCS2215, at $z=1.46$ for the first time in {\it three dimensions} by
using a set of narrowband images targeting nebular emissions from the galaxies
associated with the galaxy cluster. 
The uniqueness of this study is a combination of NB912 and NB921
narrowband filters installed in Suprime-Cam on the Subaru Telescope
which is very effective to estimate the accurate redshifts of the
star-forming galaxies at $z\sim1.46$. 
The two overlapping filters, with a slight difference in central
wavelength corresponding to $\Delta v \sim 2000$ km s$^{-1}$, cause a
difference in measured flux depending on the wavelength where the
emission line is observed in each filter. 
Both narrowband filters can detect \oii\ emission
line from galaxies at $z\sim1.46$. Thus, the ratio of \oii\ flux
measured with NB912 filter to that with NB921 filter is a good
estimator of redshift for \oii\ emission line galaxies. 

We have 41 spectroscopic redshifts for \oii\ emitters identified by
both NB912 and NB921 filters, which are mainly obtained by our
near-infrared spectroscopic follow-up with MOIRCS and FMOS on the
Subaru. We confirm that such \oii\ emitters have a ratio of
NB912-measured \oii\ flux to NB921 which is in good agreement with
that expected from the filter response functions
(Figure~\ref{fig;fluxratio}). Comparing with the spectroscopic
redshift, $z_{\rm spec}$, we show that the redshift from the
difference of line flux, $z_{\rm NB}$, can be derived with accuracy of
$\sigma((z_{\rm NB}-z_{\rm spec})/(1+z_{\rm spec}))=0.002$. 
This demonstrates that the estimation of redshift from the
emission fluxes measured with different but adjacent narrowband
filters is quite reliable. 

In this paper, we use two samples of \oii\ emission line galaxies at
$z\sim1.46$. One is a sample of 380 \oii\ emitters identified by NB912
which is constructed by \citet{hayashi2011}, while the other is a
sample of 429 \oii\ emitters identified using NB921 which is newly
constructed in this paper. Among these two samples, 170 galaxies are
found in both datasets. We then estimate the redshifts for the 170
\oii\ emitters from the fluxes measured with NB912 and NB921 and so
map the three dimensional distribution of these galaxies around the
galaxy cluster. In addition, we have a sample of \ha\ emitters at
$z=1.46$ in the field, although it is limited to the central
13.65\arcmin$\times$13.65\arcmin\ region. Due to the shallowness of 
the \ha\ data, only nine \ha\ emitters are detected. About half of them
are identified by NB921 \oii\ emitters as well. 

Our observations identify three-dimensional structures extending
towards the centre of the galaxy cluster, some of which intersect
other structures. The view we have found is consistent with a picture
of hierarchical formation in which galaxy clusters are formed
at the intersection of filamentary structures and grow through
accretion of galaxies along the large scale structures. As we have
already argued in \citet{hayashi2010,hayashi2011}, we reiterate that
this galaxy cluster is active and 
galaxies in the surrounding region
constitute the filamentary structures penetrating towards the centre
of galaxy cluster at $z=1.46$ within which the vigorous evolution of
galaxies is occurring. 

We also show the mild correlation between the local density calculated
from the three dimensional distribution and the surface density
estimated from the projected celestial distribution used in
\citet{hayashi2011}, suggesting the importance of the 3D view of 
structures for properly estimating the environment. However, we have
found results consistent with \citet{hayashi2011} in that galaxies
with a red colour tend to be located in higher density region and that
star-formation activity is not strongly dependent on the environment
within the cluster or the surrounding structures. This implies that the
growth of galaxies is on-going along the filaments and in the centre
of the galaxy cluster, and the process which quenches their activity
may work more efficiently in the core region than in the infall regions. 
Unfortunately, spectroscopic data currently available are too sparse
to discuss the change of galaxy properties in detail along the structures.   
We therefore require spectroscopic data covering multiple nebular
emission lines in the rest-frame optical to further investigate the galaxy
properties and processes critical for the evolution of cluster
galaxies. 

We conclude that the use of two adjacent narrowband filters
provides us with the accurate redshifts and emission line fluxes of
galaxies, which enables to draw a three dimensional view of galaxy
distribution as well as to properly derive the star-formation activity
of the galaxies in the whole field surveyed. In the near future, it is
expected that much larger samples of galaxies at high redshifts will
be discovered by wider-field imaging surveys such as one with Hyper
Suprime-Cam on the Subaru Telescope. These samples will be too
numerous to be observed spectroscopically. In this case, the method we
apply in this paper can be a powerful tool for the accurate estimation
of the redshift and emission line flux of star-forming galaxies. 

\section*{Acknowledgments}
We would like to thank an anonymous referee for carefully reading our
manuscript and providing helpful comments.  
Many of the data used in this paper are collected at Subaru Telescope,
which is operated by the National Astronomical Observatory of Japan.  
We thank the Subaru Telescope staff for their invaluable effort in our
observation with Suprime-Cam in service programme. 
The United Kingdom Infrared Telescope is operated by the Joint
Astronomy Centre on behalf of the Science and Technology Facilities
Council of the U.K.  
T.K. acknowledges the financial support in part by Grant-in-Aid for
the Scientific Research (Nos.\, 21340045, 24244015) by the Japanese
Ministry of Education, Culture, Sports, Science and Technology.
DS acknowledges financial support from the Netherlands Organisation
for Scientific research (NWO) through a Veni fellowship. 
IRS acknowledges support from STFC (ST/I001573/1), the ERC Advanced
Investigator programme DUSTYGAL 321334 and a Royal Society/Wolfson
Merit Award.

\bsp

\label{lastpage}


\begin{thebibliography}{}

\bibitem[\protect\citeauthoryear{{Bassett} et~al.,}{{Bassett}
  et~al.}{2013}]{Bassett2013}
{Bassett} R.  et~al., 2013, \apj, 770, 58

\bibitem[\protect\citeauthoryear{{Bertin} \& {Arnouts}}{{Bertin} \&
  {Arnouts}}{1996}]{bertin1996}
{Bertin} E.,  {Arnouts} S.,  1996, A\&AS, 117, 393

\bibitem[\protect\citeauthoryear{{Best} et~al.,}{{Best}
  et~al.}{2010}]{Best2010}
{Best} P.  et~al., 2010, arXiv:1003.5183

\bibitem[\protect\citeauthoryear{{Blakeslee} et~al.,}{{Blakeslee}
  et~al.}{2006}]{Blakeslee2006}
{Blakeslee} J.~P.  et~al., 2006, \apj, 644, 30

\bibitem[\protect\citeauthoryear{{Brammer} et~al.,}{{Brammer}
  et~al.}{2011}]{Brammer2011}
{Brammer} G.~B.  et~al., 2011, \apj, 739, 24

\bibitem[\protect\citeauthoryear{{Brodwin} et~al.,}{{Brodwin}
  et~al.}{2013}]{Brodwin2013}
{Brodwin} M.  et~al., 2013, \apj, 779, 138

\bibitem[\protect\citeauthoryear{{Cardelli}, {Clayton} \& {Mathis}}{{Cardelli}
  et~al.}{1989}]{cardelli1989}
{Cardelli} J.~A.,  {Clayton} G.~C.,    {Mathis} J.~S.,  1989, ApJ, 345, 245

\bibitem[\protect\citeauthoryear{{Casali} et~al.,}{{Casali}
  et~al.}{2007}]{casali2007}
{Casali} M.  et~al., 2007, A\&A, 467, 777

\bibitem[\protect\citeauthoryear{{Colless} et~al.,}{{Colless}
  et~al.}{2001}]{Colless2001}
{Colless} M.  et~al., 2001, \mnras, 328, 1039

\bibitem[\protect\citeauthoryear{{Daddi}, {Cimatti}, {Renzini}, {Fontana},
  {Mignoli}, {Pozzetti}, {Tozzi} \& {Zamorani}}{{Daddi}
  et~al.}{2004}]{daddi2004}
{Daddi} E.,  {Cimatti} A.,  {Renzini} A.,  {Fontana} A.,  {Mignoli} M.,
  {Pozzetti} L.,  {Tozzi} P.,    {Zamorani} G.,  2004, ApJ, 617, 746

\bibitem[\protect\citeauthoryear{{Erb}, {Bogosavljevi{\'c}} \& {Steidel}}{{Erb}
  et~al.}{2011}]{Erb2011}
{Erb} D.~K.,  {Bogosavljevi{\'c}} M.,    {Steidel} C.~C.,  2011, \apjl, 740,
  L31

\bibitem[\protect\citeauthoryear{{Faloon} et~al.,}{{Faloon}
  et~al.}{2013}]{Faloon2013}
{Faloon} A.~J.  et~al., 2013, \apj, 768, 104

\bibitem[\protect\citeauthoryear{{Gal}, {Lemaux}, {Lubin}, {Kocevski} \&
  {Squires}}{{Gal} et~al.}{2008}]{gal2008}
{Gal} R.~R.,  {Lemaux} B.~C.,  {Lubin} L.~M.,  {Kocevski} D.,    {Squires}
  G.~K.,  2008, \apj, 684, 933

\bibitem[\protect\citeauthoryear{{Galametz} et~al.,}{{Galametz}
  et~al.}{2013}]{Galametz2013}
{Galametz} A.  et~al., 2013, \aap, 559, A2

\bibitem[\protect\citeauthoryear{{Geach}, {Smail}, {Best}, {Kurk}, {Casali},
  {Ivison} \& {Coppin}}{{Geach} et~al.}{2008}]{Geach2008}
{Geach} J.~E.,  {Smail} I.,  {Best} P.~N.,  {Kurk} J.,  {Casali} M.,  {Ivison}
  R.~J.,    {Coppin} K.,  2008, \mnras, 388, 1473

\bibitem[\protect\citeauthoryear{{Gerke} et~al.,}{{Gerke}
  et~al.}{2007}]{Gerke2007}
{Gerke} B.~F.  et~al., 2007, \mnras, 376, 1425

\bibitem[\protect\citeauthoryear{{Hayashi}, {Kodama}, {Koyama}, {Tadaki} \&
  {Tanaka}}{{Hayashi} et~al.}{2011}]{hayashi2011}
{Hayashi} M.,  {Kodama} T.,  {Koyama} Y.,  {Tadaki} K.-I.,    {Tanaka} I.,
  2011, \mnras, 415, 2670

\bibitem[\protect\citeauthoryear{{Hayashi}, {Kodama}, {Koyama}, {Tanaka},
  {Shimasaku} \& {Okamura}}{{Hayashi} et~al.}{2010}]{hayashi2010}
{Hayashi} M.,  {Kodama} T.,  {Koyama} Y.,  {Tanaka} I.,  {Shimasaku} K.,
  {Okamura} S.,  2010, \mnras, 402, 1980

\bibitem[\protect\citeauthoryear{{Hayashi}, {Kodama}, {Tadaki}, {Koyama} \&
  {Tanaka}}{{Hayashi} et~al.}{2012}]{hayashi2012}
{Hayashi} M.,  {Kodama} T.,  {Tadaki} K.-I.,  {Koyama} Y.,    {Tanaka} I.,
  2012, \apj, 757, 15

\bibitem[\protect\citeauthoryear{{Hayashi}, {Sobral}, {Best}, {Smail} \&
  {Kodama}}{{Hayashi} et~al.}{2013}]{hayashi2013}
{Hayashi} M.,  {Sobral} D.,  {Best} P.~N.,  {Smail} I.,    {Kodama} T.,  2013,
  \mnras, 430, 1042

\bibitem[\protect\citeauthoryear{{Henry}, {Aoki}, {Finoguenov}, {Fotopoulou},
  {Hasinger}, {salvato}, {Suh} \& {Tanaka}}{{Henry} et~al.}{2014}]{Henry2014}
{Henry} J.~P.,  {Aoki} K.,  {Finoguenov} A.,  {Fotopoulou} S.,  {Hasinger} G.,
  {salvato} M.,  {Suh} H.,    {Tanaka} M.,  2014, \apj, 780, 58

\bibitem[\protect\citeauthoryear{{Hilton} et~al.,}{{Hilton}
  et~al.}{2010}]{hilton2010}
{Hilton} M.  et~al., 2010, ApJ, 718, 133

\bibitem[\protect\citeauthoryear{{Holden} et~al.,}{{Holden}
  et~al.}{2007}]{Holden2007}
{Holden} B.~P.  et~al., 2007, \apj, 670, 190

\bibitem[\protect\citeauthoryear{{Ichikawa} et~al.,}{{Ichikawa}
  et~al.}{2006}]{ichikawa2006}
{Ichikawa} T.  et~al., 2006, in Society of Photo-Optical Instrumentation
  Engineers (SPIE) Conference Series.

\bibitem[\protect\citeauthoryear{{Ilbert} et~al.,}{{Ilbert}
  et~al.}{2009}]{ilbert2009}
{Ilbert} O.  et~al., 2009, \apj, 690, 1236

\bibitem[\protect\citeauthoryear{{Ishiyama} et~al.,}{{Ishiyama}
  et~al.}{2013}]{ishiyama2013}
{Ishiyama} T.  et~al., 2013, \apj, 767, 146

\bibitem[\protect\citeauthoryear{{Iwamuro} et~al.,}{{Iwamuro}
  et~al.}{2012}]{iwamuro2012}
{Iwamuro} F.  et~al., 2012, \pasj, 64, 59

\bibitem[\protect\citeauthoryear{{Kennicutt}
  Jr.}{{Kennicutt}}{1998}]{kennicutt1998}
{Kennicutt} Jr. R.~C.,  1998, ARA\&A, 36, 189

\bibitem[\protect\citeauthoryear{{Kodama}, {Hayashi}, {Koyama}, {Tadaki},
  {Tanaka} \& {Shimakawa}}{{Kodama} et~al.}{2013}]{kodama2013}
{Kodama} T.,  {Hayashi} M.,  {Koyama} Y.,  {Tadaki} K.-I.,  {Tanaka} I.,
  {Shimakawa} R.,  2013, in {Thomas} D.,  {Pasquali} A.,   {Ferreras} I.,  eds,
   IAU Symposium Vol. 295, IAU Symposium. pp 74--77

\bibitem[\protect\citeauthoryear{{Kodama} et~al.,}{{Kodama}
  et~al.}{2005}]{kodama2005}
{Kodama} T.  et~al., 2005, PASJ, 57, 309

\bibitem[\protect\citeauthoryear{{Kong} et~al.,}{{Kong}
  et~al.}{2006}]{kong2006}
{Kong} X.  et~al., 2006, \apj, 638, 72

\bibitem[\protect\citeauthoryear{{Koyama}, {Kodama}, {Shimasaku}, {Hayashi},
  {Okamura}, {Tanaka} \& {Tokoku}}{{Koyama} et~al.}{2010}]{koyama2010}
{Koyama} Y.,  {Kodama} T.,  {Shimasaku} K.,  {Hayashi} M.,  {Okamura} S.,
  {Tanaka} I.,    {Tokoku} C.,  2010, MNRAS, 403, 1611

\bibitem[\protect\citeauthoryear{{Koyama}, {Kodama}, {Tadaki}, {Hayashi},
  {Tanaka}, {Smail}, {Tanaka} \& {Kurk}}{{Koyama} et~al.}{2013}]{koyama2013a}
{Koyama} Y.,  {Kodama} T.,  {Tadaki} K.-I.,  {Hayashi} M.,  {Tanaka} M.,
  {Smail} I.,  {Tanaka} I.,    {Kurk} J.,  2013, \mnras, 428, 1551

\bibitem[\protect\citeauthoryear{{Koyama} et~al.,}{{Koyama}
  et~al.}{2013}]{koyama2013b}
{Koyama} Y.  et~al., 2013, \mnras, 434, 423

\bibitem[\protect\citeauthoryear{{Lemaux} et~al.,}{{Lemaux}
  et~al.}{2012}]{Lemaux2012}
{Lemaux} B.~C.  et~al., 2012, \apj, 745, 106

\bibitem[\protect\citeauthoryear{{Miyazaki} et~al.,}{{Miyazaki}
  et~al.}{2002}]{miyazaki2002}
{Miyazaki} S.  et~al., 2002, PASJ, 54, 833

\bibitem[\protect\citeauthoryear{{Moran}, {Ellis}, {Treu}, {Smith}, {Rich} \&
  {Smail}}{{Moran} et~al.}{2007}]{Moran2007}
{Moran} S.~M.,  {Ellis} R.~S.,  {Treu} T.,  {Smith} G.~P.,  {Rich} R.~M.,
  {Smail} I.,  2007, \apj, 671, 1503

\bibitem[\protect\citeauthoryear{{Moustakas} et~al.,}{{Moustakas}
  et~al.}{2013}]{Moustakas2013}
{Moustakas} J.  et~al., 2013, \apj, 767, 50

\bibitem[\protect\citeauthoryear{{Muzzin} et~al.,}{{Muzzin}
  et~al.}{2013}]{Muzzin2013}
{Muzzin} A.  et~al., 2013, \apj, 777, 18

\bibitem[\protect\citeauthoryear{{Nakata} et~al.,}{{Nakata}
  et~al.}{2005}]{Nakata2005}
{Nakata} F.  et~al., 2005, \mnras, 357, 1357

\bibitem[\protect\citeauthoryear{{Newman}, {Ellis}, {Andreon}, {Treu},
  {Raichoor} \& {Trinchieri}}{{Newman} et~al.}{2013}]{Newman2013}
{Newman} A.~B.,  {Ellis} R.~S.,  {Andreon} S.,  {Treu} T.,  {Raichoor} A.,
  {Trinchieri} G.,  2013, arXiv: 1310.6754

\bibitem[\protect\citeauthoryear{{Ouchi} et~al.,}{{Ouchi}
  et~al.}{2004}]{ouchi2004}
{Ouchi} M.  et~al., 2004, ApJ, 611, 660

\bibitem[\protect\citeauthoryear{{Patel}, {Holden}, {Kelson}, {Illingworth} \&
  {Franx}}{{Patel} et~al.}{2009}]{Patel2009}
{Patel} S.~G.,  {Holden} B.~P.,  {Kelson} D.~D.,  {Illingworth} G.~D.,
  {Franx} M.,  2009, \apjl, 705, L67

\bibitem[\protect\citeauthoryear{{Patel}, {Kelson}, {Holden}, {Franx} \&
  {Illingworth}}{{Patel} et~al.}{2011}]{Patel2011}
{Patel} S.~G.,  {Kelson} D.~D.,  {Holden} B.~P.,  {Franx} M.,    {Illingworth}
  G.~D.,  2011, \apj, 735, 53

\bibitem[\protect\citeauthoryear{{Rodriguez Del Pino}, {Bamford},
  {Aragon-Salamanca}, {Milvang-Jensen}, {Merrifield} \& {Balcells}}{{Rodriguez
  Del Pino} et~al.}{2013}]{RodriguezDelPino2013}
{Rodriguez Del Pino} B.,  {Bamford} S.~P.,  {Aragon-Salamanca} A.,
  {Milvang-Jensen} B.,  {Merrifield} M.~R.,    {Balcells} M.,  2013, arXiv:
  1311.2842

\bibitem[\protect\citeauthoryear{{Salpeter}}{{Salpeter}}{1955}]{Salpeter1955}
{Salpeter} E.~E.,  1955, \apj, 121, 161

\bibitem[\protect\citeauthoryear{{Schlegel}, {Finkbeiner} \&
  {Davis}}{{Schlegel} et~al.}{1998}]{schlegel1998}
{Schlegel} D.~J.,  {Finkbeiner} D.~P.,    {Davis} M.,  1998, ApJ, 500, 525

\bibitem[\protect\citeauthoryear{{Shimasaku} et~al.,}{{Shimasaku}
  et~al.}{2004}]{shimasaku2004}
{Shimasaku} K.  et~al., 2004, \apjl, 605, L93

\bibitem[\protect\citeauthoryear{{Sobral} et~al.,}{{Sobral}
  et~al.}{2009}]{Sobral2009}
{Sobral} D.  et~al., 2009, \mnras, 398, 75

\bibitem[\protect\citeauthoryear{{Sobral}, {Best}, {Matsuda}, {Smail}, {Geach}
  \& {Cirasuolo}}{{Sobral} et~al.}{2012}]{Sobral2012}
{Sobral} D.,  {Best} P.~N.,  {Matsuda} Y.,  {Smail} I.,  {Geach} J.~E.,
  {Cirasuolo} M.,  2012, \mnras, 420, 1926

\bibitem[\protect\citeauthoryear{{Sobral}, {Best}, {Smail}, {Geach},
  {Cirasuolo}, {Garn} \& {Dalton}}{{Sobral} et~al.}{2011}]{Sobral2011}
{Sobral} D.,  {Best} P.~N.,  {Smail} I.,  {Geach} J.~E.,  {Cirasuolo} M.,
  {Garn} T.,    {Dalton} G.~B.,  2011, \mnras, 411, 675

\bibitem[\protect\citeauthoryear{{Sobral}, {Smail}, {Best}, {Geach}, {Matsuda},
  {Stott}, {Cirasuolo} \& {Kurk}}{{Sobral} et~al.}{2013}]{Sobral2013}
{Sobral} D.,  {Smail} I.,  {Best} P.~N.,  {Geach} J.~E.,  {Matsuda} Y.,
  {Stott} J.~P.,  {Cirasuolo} M.,    {Kurk} J.,  2013, \mnras, 428, 1128

\bibitem[\protect\citeauthoryear{{Springel} et~al.,}{{Springel}
  et~al.}{2005}]{Springel2005}
{Springel} V.  et~al., 2005, Nature, 435, 629

\bibitem[\protect\citeauthoryear{{Stanford} et~al.,}{{Stanford}
  et~al.}{2006}]{stanford2006}
{Stanford} S.~A.  et~al., 2006, ApJL, 646, L13

\bibitem[\protect\citeauthoryear{{Stott} et~al.,}{{Stott}
  et~al.}{2013}]{Stott2013}
{Stott} J.~P.  et~al., 2013, \mnras

\bibitem[\protect\citeauthoryear{{Strazzullo} et~al.,}{{Strazzullo}
  et~al.}{2013}]{Strazzullo2013}
{Strazzullo} V.  et~al., 2013, \apj, 772, 118

\bibitem[\protect\citeauthoryear{{Suzuki} et~al.,}{{Suzuki}
  et~al.}{2008}]{suzuki2008}
{Suzuki} R.  et~al., 2008, PASJ, 60, 1347

\bibitem[\protect\citeauthoryear{{Tadaki} et~al.,}{{Tadaki}
  et~al.}{2012}]{tadaki2012}
{Tadaki} K.-I.  et~al., 2012, \mnras, 423, 2617

\bibitem[\protect\citeauthoryear{{Tanaka} et~al.,}{{Tanaka}
  et~al.}{2011}]{tanaka.i2011}
{Tanaka} I.  et~al., 2011, \pasj, 63, 415

\bibitem[\protect\citeauthoryear{{Tanaka}, {Kodama}, {Arimoto}, {Okamura},
  {Umetsu}, {Shimasaku}, {Tanaka} \& {Yamada}}{{Tanaka}
  et~al.}{2005}]{Tanaka2005}
{Tanaka} M.,  {Kodama} T.,  {Arimoto} N.,  {Okamura} S.,  {Umetsu} K.,
  {Shimasaku} K.,  {Tanaka} I.,    {Yamada} T.,  2005, MNRAS, 362, 268

\bibitem[\protect\citeauthoryear{{Tanaka}, {Lidman}, {Bower}, {Demarco},
  {Finoguenov}, {Kodama}, {Nakata} \& {Rosati}}{{Tanaka}
  et~al.}{2009}]{tanaka2009}
{Tanaka} M.,  {Lidman} C.,  {Bower} R.~G.,  {Demarco} R.,  {Finoguenov} A.,
  {Kodama} T.,  {Nakata} F.,    {Rosati} P.,  2009, A\&A, 507, 671

\bibitem[\protect\citeauthoryear{{Tanaka} et~al.,}{{Tanaka}
  et~al.}{2013}]{Tanaka2013}
{Tanaka} M.  et~al., 2013, \apj, 772, 113

\bibitem[\protect\citeauthoryear{{Thomas}, {Maraston}, {Schawinski}, {Sarzi} \&
  {Silk}}{{Thomas} et~al.}{2010}]{Thomas2010}
{Thomas} D.,  {Maraston} C.,  {Schawinski} K.,  {Sarzi} M.,    {Silk} J.,
  2010, \mnras, 404, 1775

\bibitem[\protect\citeauthoryear{{Tomczak} et~al.,}{{Tomczak}
  et~al.}{2013}]{Tomczak2013}
{Tomczak} A.~R.  et~al., 2013, arXiv:1309.5972

\bibitem[\protect\citeauthoryear{{Treu}, {Ellis}, {Kneib}, {Dressler}, {Smail},
  {Czoske}, {Oemler} \& {Natarajan}}{{Treu} et~al.}{2003}]{Treu2003}
{Treu} T.,  {Ellis} R.~S.,  {Kneib} J.-P.,  {Dressler} A.,  {Smail} I.,
  {Czoske} O.,  {Oemler} A.,    {Natarajan} P.,  2003, \apj, 591, 53

\bibitem[\protect\citeauthoryear{{van den Bosch}, {Aquino}, {Yang}, {Mo},
  {Pasquali}, {McIntosh}, {Weinmann} \& {Kang}}{{van den Bosch}
  et~al.}{2008}]{vandenBosch2008}
{van den Bosch} F.~C.,  {Aquino} D.,  {Yang} X.,  {Mo} H.~J.,  {Pasquali} A.,
  {McIntosh} D.~H.,  {Weinmann} S.~M.,    {Kang} X.,  2008, \mnras, 387, 79

\bibitem[\protect\citeauthoryear{{van der Wel} et~al.,}{{van der Wel}
  et~al.}{2007}]{vanderWel2007}
{van der Wel} A.  et~al., 2007, \apj, 670, 206

\bibitem[\protect\citeauthoryear{{Yabe} et~al.,}{{Yabe}
  et~al.}{2012}]{yabe2012}
{Yabe} K.  et~al., 2012, \pasj, 64, 60

\bibitem[\protect\citeauthoryear{{Yagi}}{{Yagi}}{2012}]{yagi2012}
{Yagi} M.,  2012, \pasp, 124, 1347

\bibitem[\protect\citeauthoryear{{Yagi}, {Kashikawa}, {Sekiguchi}, {Doi},
  {Yasuda}, {Shimasaku} \& {Okamura}}{{Yagi} et~al.}{2002}]{yagi2002}
{Yagi} M.,  {Kashikawa} N.,  {Sekiguchi} M.,  {Doi} M.,  {Yasuda} N.,
  {Shimasaku} K.,    {Okamura} S.,  2002, AJ, 123, 66

\bibitem[\protect\citeauthoryear{{Yamada}, {Nakamura}, {Matsuda}, {Hayashino},
  {Yamauchi}, {Morimoto}, {Kousai} \& {Umemura}}{{Yamada}
  et~al.}{2012}]{Yamada2012}
{Yamada} T.,  {Nakamura} Y.,  {Matsuda} Y.,  {Hayashino} T.,  {Yamauchi} R.,
  {Morimoto} N.,  {Kousai} K.,    {Umemura} M.,  2012, \aj, 143, 79

\bibitem[\protect\citeauthoryear{{York} et~al.,}{{York}
  et~al.}{2000}]{York2000}
{York} D.~G.  et~al., 2000, \aj, 120, 1579

\end{thebibliography}
\end{document}